\documentclass[a4papers]{cas-sc}

\usepackage{lineno,hyperref}
\usepackage{color}
\usepackage{multirow}
\usepackage[authoryear,longnamesfirst]{natbib}
\modulolinenumbers[5]

\def\tsc#1{\csdef{#1}{\textsc{\lowercase{#1}}\xspace}}
\tsc{WGM}
\tsc{QE}

\begin{document}
\let\WriteBookmarks\relax
\def\floatpagepagefraction{1}
\def\textpagefraction{.001}

\shorttitle{
Two-component jet model for VHE gamma-ray bursts
}   

\shortauthors{Sato et al. 2022}

%\begin{frontmatter}

\title[mode = title]{
Synchrotron Self-Compton Emission in the Two-Component Jet Model for Gamma-Ray Bursts
}

\author[1]{Yuri~Sato}
\ead{yuris@phys.aoyama.ac.jp}

\author[1]{Kaori~Obayashi}

\author[2]{B.~Theodre~Zhang}

\author[1,3]{Shuta~J.~Tanaka}

\author[4,5,6,2]{Kohta~Murase}

\author[7]{Yutaka~Ohira}

\author[1,8]{Ryo~Yamazaki}

\address[1]{Department of Physical Sciences, Aoyama Gakuin University,
5-10-1 Fuchinobe, Sagamihara, Kanagawa 252-5258, Japan}
\address[2]{Center for Gravitational Physics, Yukawa Institute for Theoretical Physics, Kyoto, Kyoto 606-8502, Japan}
\address[3]{Graduate School of Engineering, Osaka University, 2-1 Yamadaoka, Suita, Osaka 565-0871, Japan}
\address[4]{Department of Physics, The Pennsylvania State University, University Park, Pennsylvania 16802, USA}
\address[5]{Department of Astronomy and Astrophysics, The Pennsylvania State University, University Park, Pennsylvania 16802, USA}
\address[6]{Center for Multimessenger Astrophysics, Institute for Gravitation and the Cosmos, The Pennsylvania State University, University Park, Pennsylvania 16802, USA}
\address[7]{Department of Earth and Planetary Science, The University of Tokyo, 7-3-1 Hongo, Bunkyo-ku, Tokyo 113-0033, Japan}
\address[8]{Institute of Laser Engineering, Osaka University,
2-6, Yamadaoka, Suita, Osaka 565-0871, Japan}

 \begin{abstract}
Gamma-ray bursts (GRBs) are intense bursts of high-energy photons (prompt emissions) caused by relativistic jets.
After the emissions, multi-wavelength afterglows, from radio to very-high-energy (VHE) gamma-ray, last for more than a few days. 
In the past three years, the VHE gamma-ray photons from four GRBs (GRBs 180720B, 190114C, 190829A and 201216C) were detected by ground-based Imaging Atmospheric Cherenkov Telescopes, such as the Major
Atmospheric Gamma Imaging Cherenkov (MAGIC) telescopes and the High Energy Stereoscopic System (H.E.S.S.).
One of them, GRB 190829A,
had some peculiar features of showing achromatic peaks in X-ray and optical bands at $1.4\times10^3$~s and being classified as low-luminosity GRBs.
Previously, we proposed a two-component jet model, which has
`narrow jet' with a small initial opening half-angle $\theta_0=0.015$~rad and large bulk Lorentz factor $\Gamma_0=350$, and `wide jet' with $\theta_0=0.1$~rad and $\Gamma_0=20$.
The narrow jet explained the early X-ray and optical emissions and apparently small isotropic gamma-ray energy and peak energy in the off-axis viewing case.
Furthermore, the late X-ray and radio (1.3 and 15.5 GHz) afterglows were emitted from the wide jet.
Here, we calculate the VHE gamma-ray flux by the synchrotron self-Compton (SSC) emission.
The multi-wavelength afterglows of GRB 190829A including the VHE gamma-ray emission are well explained by our two-component jet model.
The afterglow emissions from our two-component jet are also consistent with the observational results of GRBs 180720B, 190114C and 201216C,
when the jets are viewed on-axis.
Furthermore, we discuss the detectability of off-axis orphan afterglows by the Cherenkov Telescope Array (CTA).

\end{abstract}
 
\ExplSyntaxOn
\keys_set:nn { stm / mktitle } { nologo }
\ExplSyntaxOff

\maketitle

%\end{frontmatter}

%%%%%%%%%%%%%%%   Introduction   %%%%%%%%%%%%%%%
%%%%%%%%%%%%%%%%%%%%%%%%%%%%%%%%%%%
\section{Introduction}

Gamma-ray bursts (GRBs) are intense bursts of high-energy (between 10 keV and 10 MeV) photons (prompt emissions), and
the duration of the emission ranges between 0.1 and 1000 seconds \citep{Piran2004}.
After the prompt emission, multi-wavelength afterglows, from radio to very-high-energy (VHE) gamma-rays, last for more than a few days \citep{Meszaros1997}. 
It is believed that gamma-ray photons are emitted by relativistic jets toward us (fireball model) \citep{Piran1999, Peer2015}.
The jet is slowed down by the interaction with the ambient interstellar matter (ISM), and subsequently the external shock is formed \citep{Meszaros1993}.
Currently, the details of 
the prompt radiation mechanisms and the central engine 
have not been well understood.

In the past three years, the VHE gamma-ray photons from four GRBs, GRB 180720B  \citep{Abdalla2019}, GRB 190114C \citep{MAGIC2019a}, GRB 190829A \citep{HESS2021} and GRB 201216C \citep{Blanch2020b}, were detected by ground-based Imaging Atmospheric Cherenkov Telescopes, such as the Major Atmospheric Gamma Imaging Cherenkov (MAGIC) telescopes and the High Energy Stereoscopic System (H.E.S.S.).
Furthermore, MAGIC has reported the VHE gamma-ray photons with $\sim3\sigma$ significance from GRBs 160821B \citep{Acciari2021} and 201015A \citep{Blanch2020a}.

The VHE gamma-ray flux from GRB 190829A was detected by H.E.S.S. with 0.2--4.0 TeV energy flux of $\sim4.0\times10^{-11}~{\rm erg~s^{-1}~cm^{-2}}$ at $t\sim2\times10^4$~s after the burst onset \citep{HESS2021}.
This event 
had
much smaller isotropic-equivalent luminosity,~$L_{\rm{iso}}=10^{49}$~erg ${\rm s^{-1}}$, than typical long GRBs, and was classified as low-luminosity GRBs \citep{Chand2020}.
The prompt emission of
GRB 190829A consists of two episodes \citep{Chand2020}.
According to \citet{Chand2020}, the first emission (Episode~1) had the isotropic gamma-ray energy $E_{\rm iso,\gamma} = 3.2\times10^{49}$~erg and the peak photon energy of the $\nu F_{\nu}$ spectrum $E_p = 120.0$~keV.
At about 40~s after Episode~1, the second brighter emission with $E_{\rm iso,\gamma} = 1.9\times10^{50} $~erg and $E_p = 10.9$~keV was observed (Episode~2).
The values of $E_{\rm iso,\gamma}$ and $E_p$ of both Episodes~1 and 2 were smaller than typical long GRBs \citep{Chand2020}.
GRB 190829A was located at a low redshift $z = 0.0785$ \citep{Valeev2019}.
The very long baseline interferometry (VLBI) observations gave the upper limits on the source size in the radio bands \citep{Salafia2021}.
Furthermore, the $3\sigma$ upper limit on the optical linear polarization of $\lesssim6\%$ was also observed at $t\sim3.0\times10^3$~s \citep{Dichiara2022}.
It was associated with a type-Ic supernova SN 20190yw \citep{Hu2021}.

The multi-wavelength afterglows of GRB 190829A, were obtained in 
the 
VHE gamma-ray \citep{HESS2021}, X-ray, optical/infrared~(IR) \citep{Chand2020} and radio bands \citep{Rhodes2020, Salafia2021, Dichiara2022}.
The X-ray and optical light curves had achromatic peaks at $1.4 \times 10^{3}$~s \citep{Chand2020}.
It is difficult for the standard afterglow model to explain such achromatic peaks \citep{sari1998}, so that
the X-ray flare model \citep{Chand2020, BTZhang2021}, 
the baryon loaded outflow model \citep{Fraija2021}, the $e^+ e^-$ dust shell model \citep{LLZhang2021} and the reverse-shock emission model \citep{Salafia2021, Dichiara2022} have been proposed.
We proposed an off-axis two-component jet model in order to explain small values of $E_{\rm iso,\gamma}$ and $E_p$, and
simultaneous afterglow
peaks in X-ray and optical bands \citep{Sato2021}.

The VHE gamma-ray afterglows from bright GRBs 180720B and 190114C could 
be explained by the 
synchrotron self-Compton (SSC) mechanism \citep[e.g.][]{Derishev2019, Fraija2019a, Fraija2019b, Fraija2019c, MAGIC2019b, Asano2020, Huang2021, Derishev2021, Yamasaki2022} 
and/or the photohadronic model  \citep{Sahu2020}.
For GRB 190829A,
the external inverse-Compton \citep{BTZhang2021} and photohadronic \citep{Sahu2022} models have been proposed in addition to the SSC emission model.

It was shown in
\citet{Sato2021} that the observational results of GRB 190829A were consistent with
the off-axis two-component jet model which consists of two uniform jets (`narrow' and `wide' jets).
The early X-ray and optical light curves of GRB 190829A could be off-axis emissions from the narrow jet with initial opening half-angle $\theta_0=0.015$~rad and initial 
bulk
Lorentz factor $\Gamma_0=350$.
The wide jet, which has $\theta_0=0.1$~rad and $\Gamma_0=20$, explained the late X-ray and radio (1.3 and 15.5 GHz) afterglows.
The hydrodynamic simulation showed that GRB jets are structured with an angular dependent energy distribution 
\citep[e.g.][]{Zhang2009,Gottlieb2020,Urrutia2022}.
The two-component jet may be the most simple approximation of `the structured-jets' with envelopes.
\citet{Sato2021} did not calculate the VHE gamma-ray afterglows of GRB 190829A. 
In this paper, we discuss our off-axis two-component jet model to explain multi-wavelength afterglows including the VHE gamma-ray flux for GRB 190829A.
In the present study, we calculate the SSC flux, using the jet parameters similar to those determined by \citet{Sato2021}.
The radio emissions in 5.5 and 99.8 GHz are also computed to explain the observed ones.

\begin{figure*}
\centering 
\includegraphics[width=1.0\textwidth]{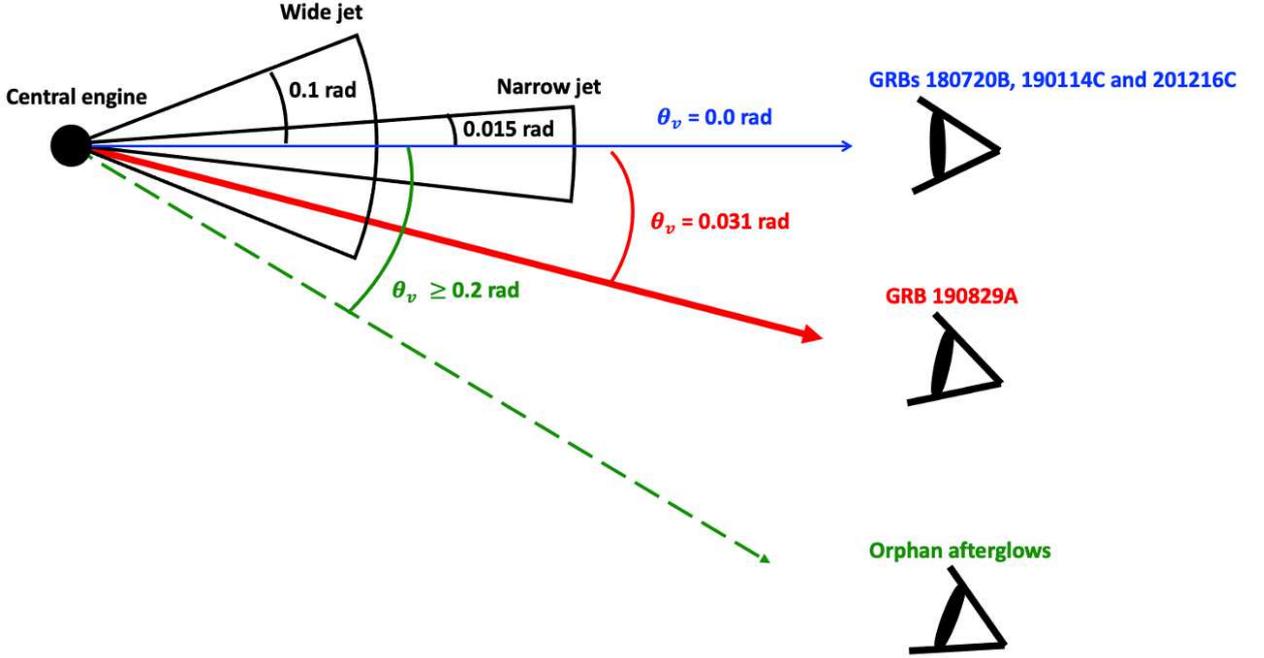}
\caption{
Schematic view of our two-component jet model for 
very-high-energy
GRBs. 
The afterglows of GRB 190829A are off-axis ($\theta_v=0.031$~rad, thick red solid arrow) emissions.
GRBs 180720B, 190114C and 201216C are observed in the case of on-axis viewing ($\theta_v=0.0$~rad, thin blue solid arrow).
Orphan afterglows are detected by viewed off-axis ($\theta_v\geq0.2$~rad, thin green dashed arrow).
}
\label{model}
\end{figure*}

Furthermore, we discuss whether our two-component jet model could explain the observational results of the other VHE
gamma-ray
events (GRBs 180720B, 190114C and 201216C).
GRB 190829A had the achromatic peaks $t\sim1.4\times10^3$~s and the small value of $E_{\rm iso,\gamma}$ and $E_p$ \citep{Chand2020}, while GRBs 180720B, 190114C and 201216C showed monotonically decaying X-ray light curve \citep{Fraija2019b, yamazaki2020} and had typical isotropic equivalent gamma-ray energy and peak energy (GRB 180720B: $E_{\rm iso,\gamma}= 6.0\times10^{53}$~erg~\citep{Abdalla2019} and $E_p=631$~keV~\citep{Duan2019}, GRB 190114C: $E_{\rm iso,\gamma} = 2.5\times10^{53}$~erg~\citep{MAGIC2019b} and $E_p = 998.6$~keV~\citep{Ravasio2019} and GRB 201216C: $E_{\rm iso,\gamma} = 4.7\times10^{53}$~erg~\citep{Blanch2020a} and $E_p = 326$~keV \citep{Huang2022}).
We explained the observational results of GRB 190829A by our off-axis two-component jet model (thick red solid arrow in Fig.~\ref{model}) \citep{Sato2021}.
The 
other
three VHE
gamma-ray
events have been considered that the jets are explained as on-axis emissions \citep[e.g.][]{Wang2019, MAGIC2019a, Asano2020, Huang2022, Rhodes2022}.
In this paper, we calculate the afterglow emissions for GRBs 180720B, 190114C and 201216C in the on-axis viewing case following previous works (thin blue solid arrow in Fig.~\ref{model}).

The afterglows which are not associated with the prompt emission are called `orphan afterglows' \citep{Piran2004}.
The gravitational lensing effect model \citep{Gao2022}, the dirty fireball model \citep{Ho2022} and the off-axis afterglow model \citep[e.g.][]{Nakar2002, Totani2002, Urata2015} have been proposed for the origin of orphan afterglows.
Orphan afterglows have not yet been definitively detected \citep[e.g.][]{Zhang2018,Ho2020,Ho2022,Huang2020}.
In this paper, we discuss 
prospects of detecting
of off-axis orphan afterglows by
the Large-Sized Telescope (LST) of
the Cherenkov Telescope Array (CTA)
(thin green dashed arrow in Fig.~\ref{model}).
Owing to  
a large field of view of CTA/LST
\citep{CTA2019}, it may have high detectability of orphan afterglows.
The detection of orphan afterglow by 
CTA/LST
may indicate the existence of the wide jet with 
the
low Lorentz factor introduced in our model.
We also discuss the detectability of multi-wavelength orphan afterglows by the extended Roentgen Survey with an Imaging Telescope Array (eROSITA), the Vera C. Rubin Legacy Survey of Space and Time (Rubin LSST), the Zwicky Transient Facility (ZTF), the Atacama Large Millimeter/sub-millimeter Array (ALMA), the next generation Very Large Array (ngVLA) and the Square Kilometre Array (SKA).

In \S~2, our numerical method is explained.
In \citet{Sato2021}, 
the number fraction of accelerated electrons 
$f_e$ was fixed to unity,
and only the Thomson limit was considered in calculating the cooling frequency and SSC flux.
In this paper, we calculate the SSC emission taking into account the Klein-Nishina effect, and 
changing the number fraction of accelerated electrons $f_e$.
In \S~3.1, we show our results of the multi-wavelength afterglow emissions for GRB 190829A.
In \S~3.2, we use our two-component jet model determined in \S~3.1 to explain the multi-wavelength light curves of GRBs 180720B, 190114C and 201216C. 
If our model with the narrow jet with large $\Gamma_0$ and the wide jet with small $\Gamma_0$ is suitable for VHE
gamma-ray
events, orphan afterglow emissions from the wide jet viewed off-axis may be observed. 
In \S~3.3, we show the detectability of off-axis orphan afterglow emissions from 
the wide jet.
Section 4 is devoted to a discussion.
In this paper, we adopt cosmological parameters $H_0 = 71~{\rm km}~{\rm s}^{-1}~{\rm Mpc}^{-1}$, $\Omega_M$ = 0.27 and $\Omega_{\Lambda}$ = 0.73 following \citet{Chand2020}.

%%%%%%%%%%%%%%%   Numerical Methods   %%%%%%%%%%%%%%%
%%%%%%%%%%%%%%%%%%%%%%%%%%%%%%%%%%%
\section{Numerical Methods}

We calculate the dynamics of a relativistic jet, which has an infinitely thin shell, following \citet{Huang2000} \citep[see][for details]{Sato2021}.
The jet propagates into the ISM with a constant number density $n_0$,
and it has an
isotropic-equivalent kinetic energy $E_{\rm{iso,K}}$,
initial 
bulk
Lorentz factor $\Gamma_0$ and
initial opening half-angle $\theta_0$.
A thin shell is formed as the jet expands.
The synchrotron radiation is computed by using a power-law electron energy distribution with an index $p$
and constant microphysics parameters $\epsilon_e$, $\epsilon_B$ and $f_e$, which are
the energy fractions of the internal energy going into power-law electrons
and magnetic field,
and the number fraction of accelerated electrons, respectively.
In \citet{Sato2021}, the number fraction of accelerated electrons $f_e$ 
was
fixed to be unity.
In this paper, we consider the case of $f_e\neq1.0$ to better explain the observed radio emissions.

In 
the
calculation of the VHE gamma-ray radiation, the SSC emission is assumed \citep{Sari2001}.
The SSC flux is affected by the Klein-Nishina effect \citep[e.g.][]{Blumenthal1970,Nakar2009, Murase2010, Murase2011, Wang2010, BTZhang2021, Jacovich2021}.
In \citet{Sato2021}, the SSC radiation is estimated in the Thomson limit,
while, in this paper, we take into account the Klein-Nishina effect using the numerical method given by \citet{BTZhang2021} 
\citep[see Appendix of ][for details]{BTZhang2021}.
We compare our numerical code with another in \citet{BTZhang2021}.
Our SSC and synchrotron cooling time is consistent with \citet{BTZhang2021} within $\sim2\%$.
Here we adopt \citet{Huang2000} in calculating the dynamics, while \citet{BTZhang2021} uses the dynamical evolution model by \citet{Nava2013},
however, the behavior of our light curve is only slightly different from \citet{BTZhang2021}.
The 
extragalactic
background light (EBL) absorption is considered in the calculation of the VHE gamma-ray flux.
We adopt the model in \citet{Franceschini2008} and use `EBL table'\footnote{https://pypi.org/project/ebltable/}.

\citet{Sato2021} proposed a two-component jet model, which consists of the narrow-fast ($\theta_0=0.015$~rad and $\Gamma_0=350$) and 
wide-slow
($\theta_0=0.1$~rad and $\Gamma_0=20$) jets.
We assume that both jets are uniform, and on co-axis, and they are emitted from the central engine at the same time.
The emissions are integrated along the equal arrival time surfaces (EATS) in calculation of the flux density $F_{\nu}$ \citep[see][for details]{Granot1999}.

%%%%%%%%%%%%%%%   Results   %%%%%%%%%%%%%%%
%%%%%%%%%%%%%%%%%%%%%%%%%%%%%%%%%%%
\section{Results of Afterglow Emission}

In this section, we show our numerical results of afterglow emissions.
In \S~3.1, the observed light curves of GRB 190829A are compared with our results.
In \S~3.2, our multi-wavelength afterglow emissions for GRBs 180720B, 190114C and 201216C are exhibited, and
we show that the VHE 
gamma-ray
events are well explained by our two-component jet with similar parameters as those for GRB 190829A.
In \S~3.3, we discuss the detectability of off-axis orphan afterglows by changing the viewing angle in order to indicate the existence of our wide jet.

%%%%%%%%%%%%%%%   GRB 190829A   %%%%%%%%%%%%%%%
%%%%%%%%%%%%%%%%%%%%%%%%%%%%%%%%%%%
\subsection{GRB 190829A}

In this sub-section, we show our numerical results of afterglow emissions in 
the
VHE gamma-ray (0.1 TeV), X-ray (5 keV), optical (V-band) and radio (1.3, 5.5, 15.5 and 99.8 GHz) bands, and together with the 
observed
data of GRB 190829A.
The VHE gamma-ray data are taken from \citet{HESS2021}.
We convert the observed energy flux in 0.2--4.0 TeV to the flux density at 0.1 TeV assuming that the photon index is 2.2 at any time \citep{HESS2021}.
The X-ray data are downloaded from the {\it Swift} team website\footnote{https://www.swift.ac.uk/xrt\_curves/00922968/} \citep{Evans2007, Evans2009}.
The observed energy flux in 0.3--10 keV is converted to the flux density in 5 keV. 
We assume that the photon index is 2.2 at any time, and it is coincident with the observed data.
The optical data are extracted from \citet{Chand2020}.
We adopt the V-band extinction $A_{\rm V} = 1.5$~mag \citep{Chand2020}.
The radio flux at 5.5 and 99.8 GHz bands and at 1.3 and 15.5 GHz bands are measured by \citet{Dichiara2022} and \citet{Rhodes2020}, respectively.

In this paper, we consider the Klein-Nishina effect in calculating the SSC process, and 
we do not fix $f_e$ to be unity.
The parameters determined by \citet{Sato2021} are somewhat modified.
Early X-ray and optical afterglow emissions are well explained by the narrow jet with
$\theta_v$ =  0.031~rad,
$\theta_0$ =  0.015~rad,
$E_{\rm{iso, K}}$ =  $4.0 \times 10^{53}$~erg,
$\Gamma_0$ =  350,
$n_0$ =  0.01~${\rm cm^{-3}}$,
$p$ =  2.44,
$\epsilon_B$ =  $6.0 \times 10^{-5}$,
$\epsilon_e$ =  $3.5 \times 10^{-2}$ and
$f_e$ =  0.2.
A two-component jet model is considered in this work, in which another `wide jet' is introduced.
The parameters of the wide jet are 
$\theta_v$ =  0.031~rad,
$\theta_0$ =  0.1~rad,
$E_{\rm{iso, K}}$ =  $1.0 \times 10^{53}$ erg,
$\Gamma_0$ =  20,
$n_0$  =  0.01~${\rm cm^{-3}}$,
$p$ =  2.2,
$\epsilon_B$ =  $1.0 \times 10^{-5}$,
$\epsilon_e$ =  0.29 and
$f_e$ =  0.35.
The superposition of each jet emission is compared with the observed light curve.

\begin{figure*}
\centering 
\includegraphics[width=0.7\textwidth]{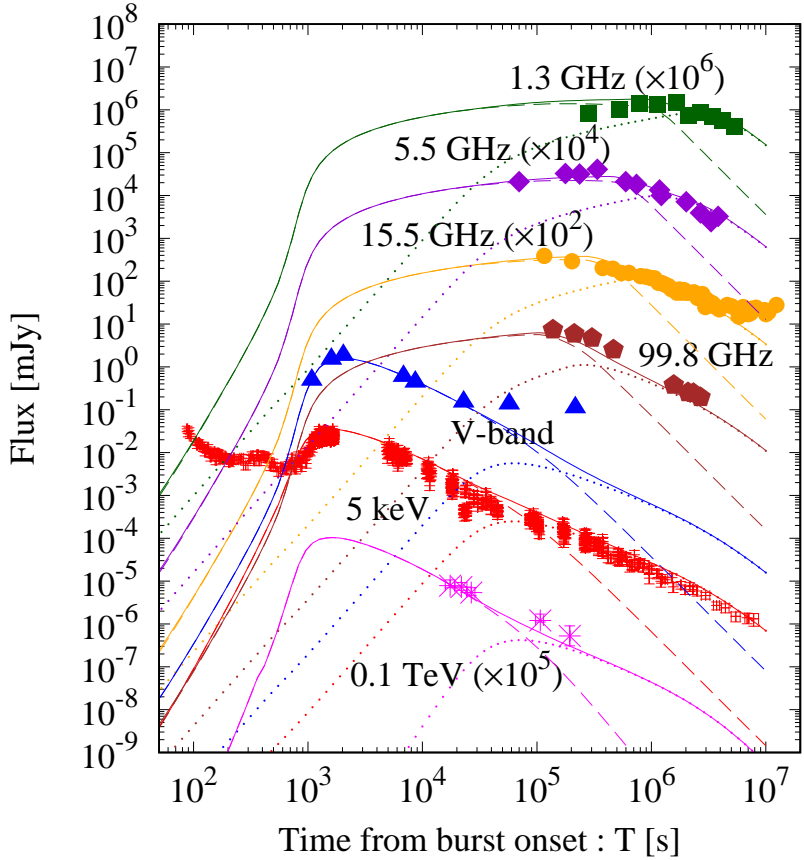}
\caption{
Observed data of GRB~190829A (VHE gamma-ray (0.1~TeV: magenta points), X-ray~(5~keV: red points), optical~(V-band: blue triangles), radio bands~(1.3~GHz: green squares, 5.5~GHz: violet diamonds, 15.5~GHz: orange filled-circles, 99.8~GHz: brown pentagons)),
together with 
multi-wavelength
afterglow modeling (VHE gamma-ray (0.1~TeV: magenta), X-ray (5~keV: red), optical (V-band: blue), radio bands (1.3~GHz: green, 5.5~GHz: violet, 5.5~GHz: orange, 99.8~GHz: brown)).
Afterglow light curves calculated by our two-component jet model -- solid lines are the sum of the narrow jet (dashed lines: $\theta_v$ = $0.031$~rad, $\theta_0$ = $0.015$~rad, $E_{\rm{iso, K}}$ = $4.0 \times 10^{53}$~erg, $\Gamma_0$ = $350$, $n_0$ = $0.01$~${\rm cm^{-3}}$, $p$ = $2.44$, $\epsilon_B$ = $6.0 \times 10^{-5}$, $\epsilon_e$ = $3.5 \times 10^{-2}$, and $f_e$ = $0.2$) and wide jet (dotted lines: $\theta_v$ = $0.031$ rad, $\theta_0$ = $0.1$~rad, $E_{\rm{iso, K}}$ = $1.0 \times 10^{53}$~erg, $\Gamma_0$ = $20$, $n_0$ = $0.01$~${\rm cm^{-3}}$, $p$ = $2.2$, $\epsilon_B$ = $1.0 \times 10^{-5}$, $\epsilon_e$ = $0.29$, and $f_e$ = $0.35$).
}
\label{lightcurve}
\end{figure*}

The numerical results of \citet{Sato2021} with $f_e=1.0$ in 1.3 and 15.5 GHz bands exceeded the observed data within about a factor of three.
We modify the value of $f_e$ from $1.0$ \citep{Sato2021} to $0.2$~and~$0.35$ for the narrow and wide jets, respectively.
The value of the typical frequency $\nu_m$ in the case of $f_e\neq1.0$ is larger than that in the case of $f_e=1.0$.
After $t\sim10^{5}$~s, the typical and absorption frequencies, $\nu_m$ and $\nu_a$, respectively, 
satisfy
$\nu_a<1.3$~GHz and 15.5~GHz~$<\nu_m$.
Then, the fluxes in 1.3 and 15.5 GHz bands depend on $f_e$ 
($F_{\nu}
\propto {f_e}^{5/3}$).
The radio fluxes are dim in the case of $f_e<1.0$.
As a result, our numerical light curves better match with the observed ones (orange and green lines in Fig.~\ref{lightcurve}).
Moreover, the late ($t\sim10^{6-7}$~s) X-ray emission calculated in \citet{Sato2021} was about a factor of two smaller than the observed flux.
The numerically calculated X-ray light curve is also improved by taking care of the Klein-Nishina effect (red lines in Fig.~\ref{lightcurve}).
When the Klein-Nishina effect is taken into account, the cooling frequency $\nu_c$ becomes larger because of less cooling compared with the Thomson limit.
Since the X-ray band exceeds $\nu_c$, the X-ray flux becomes brighter.

The VHE gamma-ray and X-ray fluxes are affected by the Klein-Nishina effect.
The observed light curves in VHE gamma-ray and X-ray bands are well explained (magenta and red lines in Fig.~\ref{lightcurve}).
We consider the 99.8 and 5.5 GHz fluxes, which had not yet 
been
reported in \citet{Sato2021}.
The narrow jet becomes trans-relativistic ($\Gamma<10$) at $t\sim 10^5$~s, and the relation among the absorption frequency $\nu_a$, the typical frequency $\nu_m$ and the cooling frequency $\nu_c$ satisfy $\nu_a<\nu_m<\nu_c$.
The typical frequency $\nu_m$ depends on time $t$, and it crosses 99.8 GHz around $1.5\times10^5$~s.
Then, the 99.8 GHz light curve has a peak.
After that, $\nu_m$ is lower than 99.8 GHz band and the flux obeys the scaling $F_{\nu}\propto t^{-p}=t^{-2.44}$ \citep{Gao2013}.
The observed data in 99.8 GHz decays slower than our numerical result.
It is hard for the narrow jet to explain the observed data (brown dashed line in Fig.~\ref{lightcurve}).
The wide jet becomes trans-relativistic around $t\sim 10^5$~s and enters the Newtonian phase at $t\sim2.0\times10^6$.
From $10^5$ to $10^7$~s, the break frequencies $\nu_a$, $\nu_m$ and $\nu_c$ satisfy the relation $\nu_a<\nu_m<\nu_c$.
The value of $\nu_m$ crosses 99.8 GHz band around $2.5\times10^5$~s, at which the flux takes 
maximum
and after that it follows as $F_{\nu}\propto t^{-3(5p-7)/10}=t^{-1.2}$ \citep{Gao2013} (brown dotted line in Fig.~\ref{lightcurve}).
The sum of the narrow and wide jet emissions is coincident with the observed 99.8 GHz afterglow (brown solid line in Fig.~\ref{lightcurve}).
Subsequently, $\nu_m$ for the narrow and wide jet emissions crosses 5.5 GHz at $8.0\times10^5$~s and $5.5\times10^6$~s, respectively, and then their light curves have maxima, after which they decay in the same way as 99.8 GHz band (violet dashed and dotted lines in Fig.~\ref{lightcurve}).
The observed 5.5 GHz light curve is explained by the sum of both components (violet solid line in Fig.~\ref{lightcurve}).

%%%%%%%%%%%%%%%   VHE evnts   %%%%%%%%%%%%%%%
%%%%%%%%%%%%%%%%%%%%%%%%%%%%%%%%%%%
\subsection{GRB 180720B, GRB 190114C and GRB 201216C}

In \S~3.1, the observational results of GRB 190829A are explained by our two-component jet model.
We use our model 
described
in \S~3.1 to show that the VHE gamma-ray (0.1 TeV), high-energy (HE) gamma-ray (0.5 GeV), 
X-ray (5 keV and 500 keV),
optical (r-band) and radio (9 and 10 GHz) afterglows of GRB 180720B, 190114C and 201216C are also well explained by similar parameters.
In particular, we fix the opening angles and initial Lorentz factors of both jets as in the previous sub-section.

The X-ray (5 keV) data of the three events are obtained from the {\it Swift} team website.
The VHE gamma-ray (0.1--0.44 TeV), HE gamma-ray (0.1--10 GeV) and optical (r-band) data of GRB 180720B are extracted from \citet{Abdalla2019}.
The observational data of GRB 190114C in 
the
VHE gamma-ray (0.2--1 TeV), HE gamma-ray (0.1--1 GeV),
hard X-ray (10--1000 keV),
optical (r-band) and radio (9 GHz) bands are taken from \citet{MAGIC2019a}.
The optical (r-band) and radio (10 GHz) data of GRB 201216C are derived from \citet{Rhodes2022}.
We assume the r-band extinction $A_r=0.8$~mag for GRBs 180720B and 190114C
, and $A_r=1.8$ mag for GRB 201216C.
Redshifts are $z=0.653,~0.4245,$ 
and
$1.1$ for GRBs 180720B \citep{Abdalla2019}, 190114C \citep{MAGIC2019a} 
and
201216C \citep{Blanch2020b}, respectively.
Adopted parameters, as well as those by previous works, are summarized in
Tables
1, 2 and 3.

\begin{figure*}
\centering
\begin{minipage}{0.3\linewidth}
\centering
\includegraphics[height=1.0\textwidth]{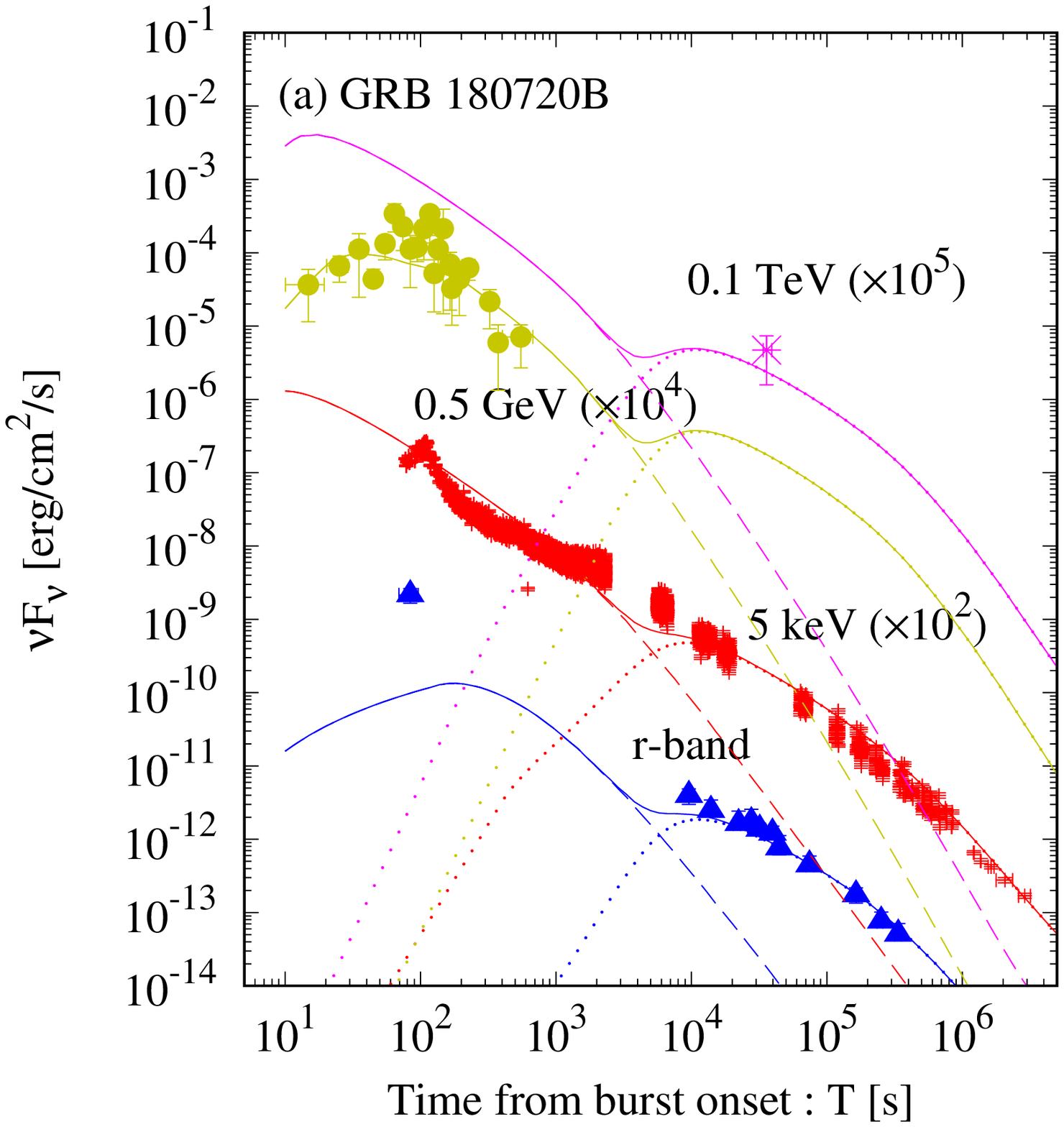}
\end{minipage}
\begin{minipage}{0.3\linewidth}
\centering
\includegraphics[height=1.0\textwidth]{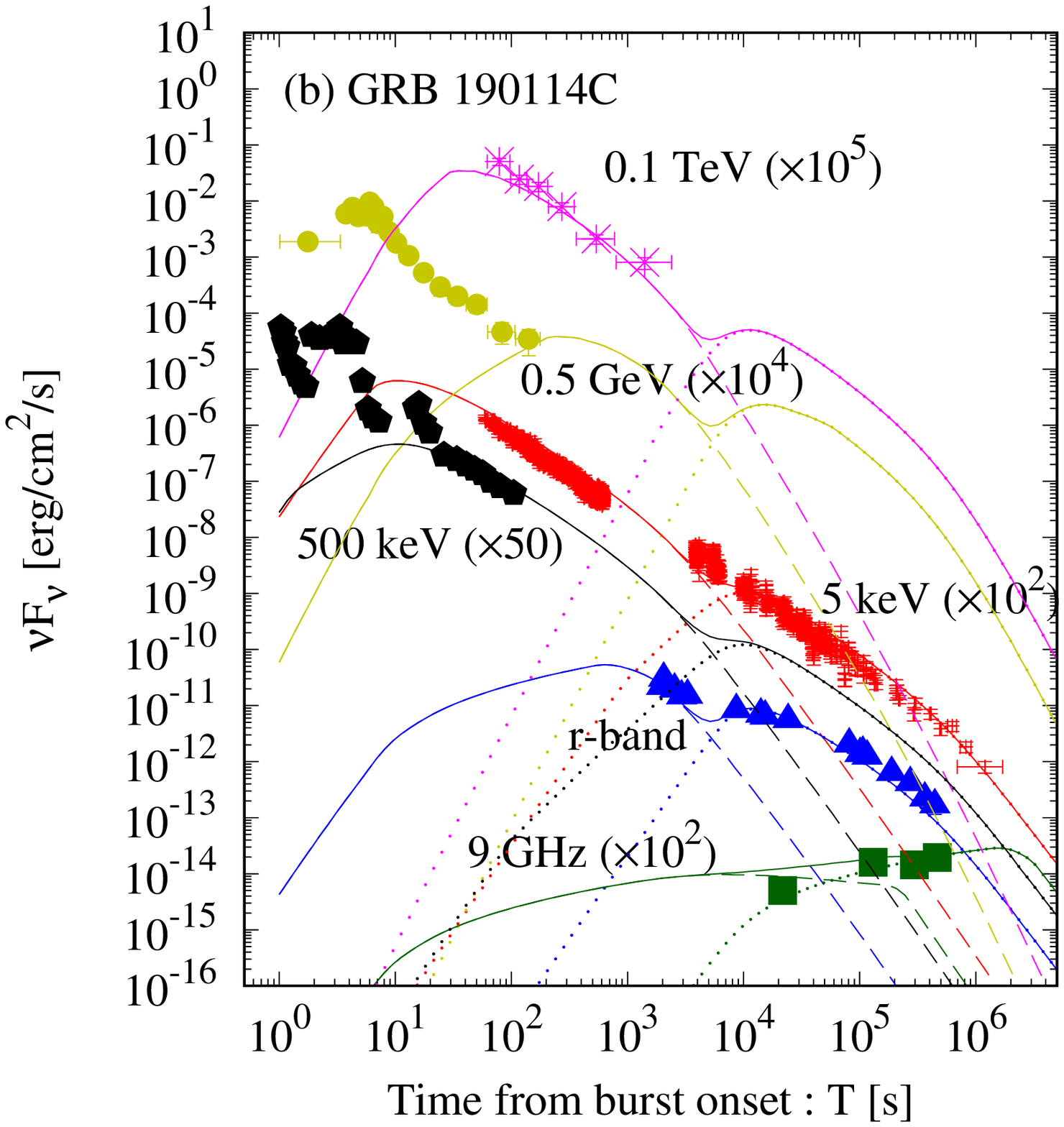}
\end{minipage}
\begin{minipage}{0.3\linewidth}
\centering
\includegraphics[height=1.0\textwidth]{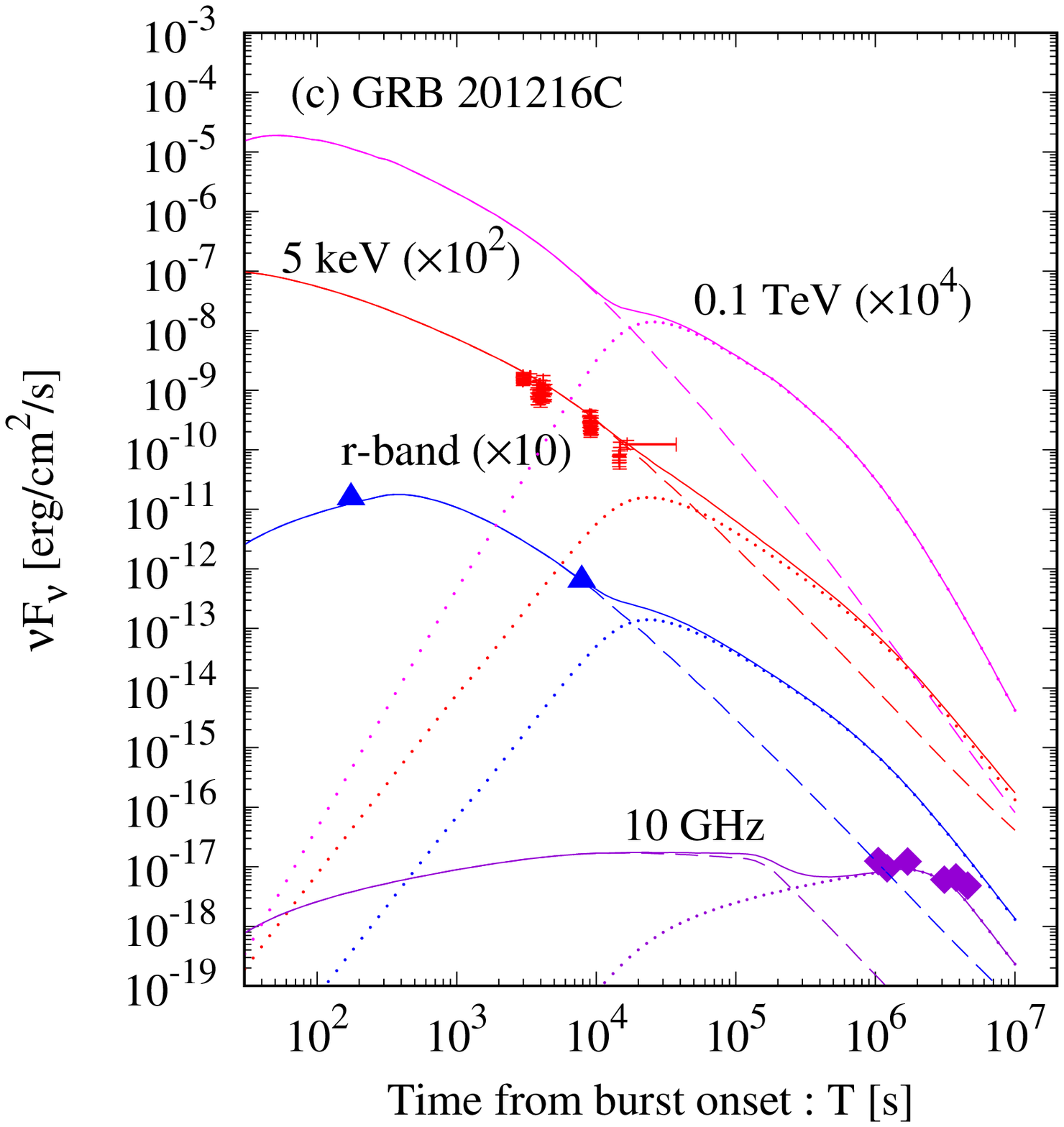}
\end{minipage}
\caption{
Observed data (VHE gamma-ray (0.1~TeV: magenta points), HE gamma-ray (0.5~GeV: yellow filled-circles),
X-ray (5~keV: red points and 
500~keV: black pentagons),
optical (r-band: blue triangles) and radio bands (9~GHz: green squares and 10~GHz: violet diamonds)) of GRB 180720B (panel (a)), 190114C (panel (b)) and 201216C (panel (c)),
together with afterglow light curves calculated by our two-component jet model -- solid lines are the sum of the narrow (dashed lines) and wide (dotted lines) jets in 
the
VHE gamma-ray (0.1~TeV: magenta), HE gamma-ray (0.5~GeV yellow), 
X-ray (5~keV: red and 500~keV: black),
optical (r-band: blue) and radio bands (1.3~GHz: green and 10~GHz: purple).
The panel (a) (GRB 180720B) shows the result of our narrow ($\theta_v$ = $0.0$~rad, $\theta_0$ = $0.015$~rad, $E_{\rm{iso, K}}$ = $4.0 \times 10^{53}$~erg, $\Gamma_0$ = $350$, $n_0$ = $10$~${\rm cm^{-3}}$, $p$ = $2.4$, $\epsilon_B$ = $5.0 \times 10^{-4}$, $\epsilon_e$ = $5.0 \times 10^{-3}$, and $f_e$ = $0.2$) and wide ($\theta_v$ = $0.0$~rad, $\theta_0$ = $0.1$~rad, $E_{\rm{iso, K}}$ = $1.0 \times 10^{53}$~erg, $\Gamma_0$ = $20$, $n_0$ = $10$~${\rm cm^{-3}}$, $p$ = $2.2$, $\epsilon_B$ = $9.0 \times 10^{-4}$, $\epsilon_e$ = $0.09$, and $f_e$ = $0.4$) jets.
In panel (b) (GRB 190114C), we show our result of the narrow ($\theta_v$ = $0.0$~rad, $\theta_0$ = $0.015$~rad, $E_{\rm{iso, K}}$ = $4.0 \times 10^{53}$~erg, $\Gamma_0$ = $350$, $n_0$ = $3$~${\rm cm^{-3}}$, $p$ = $2.8$, $\epsilon_B$ = $6.0 \times 10^{-5}$, $\epsilon_e$ = $9.0 \times 10^{-3}$, and $f_e$ = $0.1$) and wide ($\theta_v$ = $0.0$~rad, $\theta_0$ = $0.1$~rad, $E_{\rm{iso, K}}$ = $1.0 \times 10^{53}$~erg, $\Gamma_0$ = $20$, $n_0$ = $3$~${\rm cm^{-3}}$, $p$ = $2.6$, $\epsilon_B$ = $9.0 \times 10^{-4}$, $\epsilon_e$ = $8.0 \times 10^{-2}$, and $f_e$ = $0.2$) jets.
In panel (c) (GRB 201216C), the sum of the narrow ($\theta_v$ = $0.0$~rad, $\theta_0$ = $0.015$~rad, $E_{\rm{iso, K}}$ = $4.0 \times 10^{53}$~erg, $\Gamma_0$ = $350$, $n_0$ = $1$~${\rm cm^{-3}}$, $p$ = $2.3$, $\epsilon_B$ = $6.0 \times 10^{-5}$, $\epsilon_e$ = $3.5 \times 10^{-2}$, and $f_e$ = $0.4$) and wide ($\theta_v$ = $0.0$~rad, $\theta_0$ = $0.1$~rad, $E_{\rm{iso, K}}$ = $1.0 \times 10^{53}$~erg, $\Gamma_0$ = $20$, $n_0$ = $1$~${\rm cm^{-3}}$, $p$ = $2.8$, $\epsilon_B$ = $5.0 \times 10^{-5}$, $\epsilon_e$ = $0.1$, and $f_e$ = $0.2$) jets are shown.
}
\label{VHE}
\end{figure*}

The three events are considered in the case of on-axis viewing ($\theta_v$ = 0.0 rad).
Unless otherwise stated, the narrow and wide jet parameters ($\theta_0$, $E_{\rm{iso, K}}$ and $\Gamma_0$) are the same as those given in Section 3.1, while
$n_0$,
$p$,
$\epsilon_B$,
$\epsilon_e$ and
$f_e$ are changed.
As seen in  Figs.~\ref{VHE}(a), (b) and (c), our two-component jet model roughly explains the observed afterglow emissions in all wavelengths in GRBs 180720B, 190114C and 201216C.

For GRB 180720B, we adopt the narrow jet with
$n_0$ = 10.0~${\rm cm^{-3}}$,
$p$ =  $2.4$,
$\epsilon_B$ = $5.0 \times 10^{-4}$,
$\epsilon_e$ = $5.0 \times 10^{-3}$ and
$f_e$ = $0.2$.
The wide jet is introduced with the parameters,
$p$ = $2.2$,
$\epsilon_B$ =  $9.0 \times 10^{-4}$,
$\epsilon_e$ =  $9.0 \times 10^{-2}$ and
$f_e$ = $0.4$.
These values of $p$, $E_{\rm{iso, K}}$ and $\epsilon_B$ of \citet{Wang2019} are almost close to ours,
while our $n_0$ is larger than \citet{Wang2019}.

For GRB 190114C, the parameters of
$n_0$ =  $3.0$~${\rm cm^{-3}}$,
$p$ =  $2.8$,
$\epsilon_B$ = $6.0 \times 10^{-5}$,
$\epsilon_e$ = $9.0 \times 10^{-3}$ and
$f_e$ =  $0.1$
are adopted for the narrow jet.
The wide jet with
$p$ =  $2.6$,
$\epsilon_B$ =  $9.0 \times 10^{-4}$,
$\epsilon_e$ =  $8.0 \times 10^{-2}$ and
$f_e$ =  $0.2$
are used.
Parameters adopted by \citet{MAGIC2019a}, 
\citet{Wang2019} 
and
\citet{Asano2020} 
are roughly similar to our parameters, $n_0$, $p$, $E_{\rm{iso, K}}$, $\epsilon_B$ and $\epsilon_e$.
In \citet{MAGIC2019a} and \citet{Wang2019}, the number fraction of accelerated electrons was adopted as $f_e=1.0$, while in \citet{Asano2020} and this paper, it is calculated with $f_e\neq1.0$.
When the values of $E_{\rm{iso, K}}/f_e$, $f_e\epsilon_B$, $f_e\epsilon_e$ and $n_0/f_e$ are same, the behaviors of afterglow light curves are similar \citep{Eichler2005}.
The SSC flux with small $\epsilon_B$ is brighter than that with large $\epsilon_B$, so that VHE
gamma-ray
events may have small $\epsilon_B$ \citep{Miceli2022}.
In particular, we compare the value of $f_e\epsilon_B$ in the case of $f_e\neq1.0$.
For our narrow and wide jet parameters, we get $f_e\epsilon_B\sim6.0\times10^{-6}$ and $\sim1.8\times10^{-4}$, respectively.
In \citet{Asano2020}, $f_e\epsilon_B\sim2.7\times10^{-4}$ was obtained.
The value of $f_e\epsilon_B$ for 
our
narrow jet is small like that of \citet{Asano2020}, while that of 
our
wide jet and \citet{Asano2020} is similar.
Previous works \citep{MAGIC2019a,Wang2019,Asano2020} supposed a single component.
In \citet{MAGIC2019a}, the theoretically calculated radio emissions are about an order of magnitude brighter than the observed ones.
\citet{Wang2019} and \citet{Asano2020} did not discuss the radio emission.
In the present work, it is found that the two-component jet model well explains the observational radio flux.

For GRB 201216C, we use 
$n_0$ =  $1.0$~${\rm cm^{-3}}$,
$p$ =  $2.3$,
$\epsilon_B$ =  $6.0 \times 10^{-5}$,
$\epsilon_e$ =  $3.5 \times 10^{-2}$ and
$f_e$ =  $0.4$
in the narrow jet.
The parameters of the wide jet are
$p$ =  $2.8$,
$\epsilon_B$ =  $5.0 \times 10^{-5}$,
$\epsilon_e$ =  $0.1$ and
$f_e$ =  $0.2$.
\citet{Huang2022} obtained parameters considering EATS
effect,
and the values of $n_0$, $p$, $E_{\rm{iso, K}}$  and $\epsilon_B$ are almost consistent with ours.
In \citet{Rhodes2022}, the jet-cocoon model was adopted.
In particular, the narrow-core component ($\Gamma_0>100$) 
explained the observed afterglow in the X-ray and optical, while
the wider low-energy component ($\Gamma_0<10$) 
was consistent with the radio emission.
The value of $E_{\rm{iso, K}}$ of our wide jet is larger than that of the cocoon component of \citet{Rhodes2022}.
The stellar wind environment was considered in \citet{Rhodes2022}, and then, the density 
in front
of the shell is
$\sim10^{3-5}$~${\rm cm^{-3}}$ at 10 days.
Then, for both \citet{Rhodes2022} and our wide jet parameter, 
the absorption frequency $\nu_a$ is larger than 10 GHz, and
the flux depends on $E_{\rm{iso, K}}$ and $n_0$ 
($F_{\nu}\propto E_{\rm{iso, K}}^{5/6}~{n_0}^{1/2}$).
The flux with large $E_{\rm{iso, K}}$ and small $n_0$ is consistent with the flux in the case of small $E_{\rm{iso, K}}$ and large $n_0$, so that
the observed radio flux of GRB 201216C is also explained by our wide jet with different values of $E_{\rm{iso, K}}$ and $n_0$ from \citet{Rhodes2022}. 

As seen in  Fig.~\ref{VHE}(a), the narrow jet explains the early time ($t\lesssim2.0\times10^3$~s) observational results of GRB 180720B in the HE
gamma-ray
and X-ray bands, while the VHE
gamma-ray,
X-ray and optical emissions from the wide jet are consistent with the observed data at the late ($t\gtrsim5.0\times10^3$~s) epoch.
For GRB 190114C, the early ($t\lesssim6.0\times10^3$~s) observed VHE
gamma-ray,
X-ray (5~keV)
and optical fluxes are consistent with the narrow jet emissions, and at the late time ($t\gtrsim6.0\times10^3$~s) the wide jet 
described
the observed light curves in the 
X-ray (5~keV)
and optical bands (see Fig.~\ref{VHE}(b)).
The early ($t\lesssim3.0\times10^4$~s) and late ($t\gtrsim10^5$~s) observational results in the radio band is 
well matched
by the narrow and wide jets, respectively.
Our narrow jet also fits with the observed data in the hard X-ray (500~keV) band ($t>10$~s).
As seen in Fig.~\ref{VHE}(c), the early ($t\lesssim2.0\times10^4$~s) X-ray and optical afterglows of GRB 201216C are explained by the narrow jet, while the late ($t\gtrsim10^6$~s) radio 
data
is consistent with the wide jet
emission.
Therefore, the observational results of the three VHE
gamma-ray
events are consistent with our two-component jet model.
For all VHE
gamma-ray
events, the wide jet is necessary.

\begin{table}
\caption{Parameters for modeling of GRB 180720B}
\label{table:180720B}
\centering
\begin{tabular}{lcccccc}
\hline
 & ~~$n_0$~${[\rm cm^{-3}]}$~~ & ~~$p$~~ & ~~$E_{\rm{iso, K}}$~[erg]~~ & ~~$\epsilon_B$~~ & ~~$\epsilon_e$~~ & ~~$f_e$~~  \\
\hline 
\multirow{2}{*}{Present work}~(narrow jet) & \multirow{2}{*}{~~$10.0$~~} & ~~$2.4$~~ & ~~$4.0\times10^{53}$~~ & ~~$5.0\times10^{-4}$~~ & ~~$5.0\times10^{-3}$~~~ & ~~~$0.2$~~~ \\
\multirow{2}{*}{~~~~~~~~~~~~~~~~} (wide jet) & & ~~$2.2$~~ & ~~$1.0\times10^{53}$~~ & ~~$9.0\times10^{-4}$~~ & ~~$9.0\times10^{-2}$~~ & ~~~$0.4$~~ \\
\citet{Wang2019} & ~~$0.1$~~ & ~~$2.4$~~ & ~~$1.0\times10^{54}$~~ & ~~$1.0\times10^{-4}$~~ & ~~$0.1$~~ & ~~$1.0$~~ \\
\hline
\end{tabular}
\end{table}

\begin{table}
\caption{Parameters for modeling of GRB 190114C}
\label{table:190114C}
\centering
\begin{tabular}{lcccccc}
\hline
 & $n_0$~${[\rm cm^{-3}]}$ & $p$ & $E_{\rm{iso, K}}$~[erg] & $\epsilon_B$ & $\epsilon_e$ & $f_e$  \\
\hline 
\multirow{2}{*}{Present work}~(narrow jet) & \multirow{2}{*}{$3.0$} & $2.8$ & $4.0\times10^{53}$ & $6.0\times10^{-5}$ & $9.0\times10^{-3}$ & $0.1$ \\
\multirow{2}{*}{~~~~~~~~~~~~~~~~} (wide jet) & & $2.6$ & $1.0\times10^{53}$ & $9.0\times10^{-4}$ & $8.0\times10^{-2}$ & $0.2$ \\
\citet{MAGIC2019a} & $0.5$--$5$ & $2.4$--$2.6$ & $\gtrsim3.0\times10^{53}$ & $(0.05$--$1.0)\times10^{-3}$ & $0.05$--$0.15$ & $1.0$ \\
\citet{Wang2019} & $0.3$ & $2.5$ & $6.0\times10^{53}$ & $4.0\times10^{-5}$ & $7.0\times10^{-2}$ & $1.0$ \\
\citet{Asano2020} & $1.0$ & $2.3$ & $1.0\times10^{54}$ & $9.0\times10^{-4}$ & $6.0\times10^{-2}$ & $0.3$ \\
\hline
\end{tabular}
\end{table}

\begin{table}
\caption{Parameters for modeling of GRB 201216C}
\label{table:201216C}
\centering
\begin{tabular}{lcccccc}
\hline
 & ~$n_0$~${[\rm cm^{-3}]}$~ & ~$p$~ & $E_{\rm{iso, K}}$~[erg] & $\epsilon_B$ & ~$\epsilon_e$~ & ~$f_e$~  \\
\hline 
\multirow{2}{*}{Present work}~(narrow jet) & \multirow{2}{*}{~$1.0$~} & ~$2.3$~ & $4.0\times10^{53}$ & $6.0\times10^{-5}$ & ~$3.5\times10^{-2}$~ & ~$0.4$~ \\
\multirow{2}{*}{~~~~~~~~~~~~~~~~} (wide jet) & & ~$2.8$~ & $1.0\times10^{53}$ & $5.0\times10^{-5}$ & ~$0.1$~ & ~$0.2$~ \\
\citet{Huang2022} (EATS) & ~$0.5$~ & ~$2.1$~ & $6.0\times10^{53}$ & $8.0\times10^{-5}$ & ~$0.6$~ & ~$1.0$~ \\
\citet{Rhodes2022} & \multirow{2}{*}{~(wind)~} & \multirow{2}{*}{~$2.0$--$2.4$~} & \multirow{2}{*}{$(0.6-10)\times10^{52}$} & \multirow{2}{*}{$(0.05$--$4.0)\times10^{-3}$} & \multirow{2}{*}{~$0.04$--$0.1$~} & \multirow{2}{*}{~$1.0$~} \\
(narrow-core component) & & & & & & \\
\citet{Rhodes2022} & \multirow{2}{*}{(wind)} & \multirow{2}{*}{~$2.0$~} & \multirow{2}{*}{$(0.02$--$50)\times10^{48}$} & \multirow{2}{*}{$1.0\times10^{-2}$} & \multirow{2}{*}{~$0.1$~} & \multirow{2}{*}{~$1.0$~} \\
(cocoon component) & & & & & & \\
\hline
\end{tabular}
\end{table}

%%%%%%%%%%%%%%%   Orphan Afterglow   %%%%%%%%%%%%%%%
%%%%%%%%%%%%%%%%%%%%%%%%%%%%%%%%%%%
\subsection{Orphan Afterglows}

In \S~3.1 and 3.2, it is shown that our two-component jet model well explains the observational results of the previous VHE
gamma-ray
events (GRBs 180720B, 190114C, 190829A and 201216C).
The existence of the wide jet is common.
If VHE
gamma-ray
events are accountable by our two-component jet, orphan afterglow emissions may be observed in the case of large off-axis viewing angle.
In this sub-section, we study the detectability of off-axis orphan afterglow emissions from our two-component jet in 
the
VHE gamma-ray, X-ray, optical and radio bands by 
CTA/LST, eROSITA, Rubin LSST, ZTF, ALMA, ngVLA and SKA, respectively.
Throughout this sub-section, the 
$g$-band
extinction is assumed as a typical value 
$A_{g}= 2.8$~mag.

The viewing angle $\theta_v$ should be larger than 0.1 rad, which corresponds to the value of the initial jet opening half-angle of the wide jet.
This is necessary because the prompt emissions from the narrow and wide jets are not detected.
Since the narrow jet has a small $\theta_0=0.015$~rad, the afterglow emissions from the narrow jet are dimmer than those from the wide jet.
In this sub-section, the afterglow fluxes from the wide jet are considered.

When we change the viewing angle, $E_{\rm iso,\gamma}(\theta_v)$ and $E_p(\theta_v)$ are estimated
using the method given by \citet{Donaghy2006}, \citet{Graziani2006} and \citet{Ioka2001} \citep[see also][]{Yamazaki2002, Yamazaki2003a, Yamazaki2003b, Yamazaki2004,Sato2021}.
Here we assume that the narrow and wide jets have the values of $E_{\rm iso,\gamma}(\theta_v=0.0$~rad) and $E_p(\theta_v=0.0$~rad) based on our model for Episode~1 and 2 of GRB 190829A
\citep{Sato2021},
that is 
$E_{\rm iso,\gamma}(\theta_v=0.0$~rad)=$2.7\times10^{53}$~erg and $E_p(\theta_v=0.0$~rad)=3.7~MeV for the narrow jet,
and $E_{\rm iso,\gamma} = 1.9\times10^{50} $~erg and $E_p = 10.9$~keV for the wide jet.
Then, we obtain
off-axis ($\theta_v=0.2$~rad) quantities are obtained as $E_{\rm iso,\gamma}(\theta_v=0.2~{\rm rad})=1.3\times10^{48}$~erg and $E_p(\theta_v=0.2~{\rm rad})=1.5$~keV for the narrow jet, and
$E_{\rm iso,\gamma}(\theta_v=0.2~{\rm rad})=9.2\times10^{47} $~erg and $E_p(\theta_v=0.2~{\rm rad})=1.9$~keV for the wide jet.
In the case of $\theta_v\geq0.2$~rad, such low-energy prompt emission may be difficult to be detected unless 
the
distance to the source is very nearby.
In this sub-section, we use the wide jet parameters ($\theta_0$ =  0.1~rad, $E_{\rm{iso, K}}$ =  $1.0 \times 10^{53}$~erg, $\Gamma_0$ = 20,~$p$ = 2.2, $\epsilon_B$ = $1.0 \times 10^{-5}$, $\epsilon_e$ =  0.29, and $f_e$ =  0.35), and change the viewing angle $\theta_v=0.0,~0.2$ and $0.3$~rad, the ISM density $n_0=0.01$ and $1.0$~${\rm cm^{-3}}$ and the redshift $z=0.05$ and $0.5$.

We consider the case of the observation with four CTA/LSTs (CTA/4~LSTs).
Although 
CTA/4~LSTs are
most sensitive around 2~TeV 
\protect\footnotemark[3],
such high-energy photons are heavily absorbed by EBL.
Here we draw the VHE gamma-ray light curves at $h \nu = 0.3$~TeV, since it is less affected by the EBL attenuation and it can be still observed with good sensitivity of 
CTA/4~LSTs.
Figure~\ref{CTA} shows the results.
In the case of  $z=0.05$ (Fig.~\ref{CTA}(a)), one can find that 
CTA/4~LSTs have
the capability to detect off-axis ($\theta_v=0.2$--0.3~rad) orphan afterglows
if the ambient density is $n_0=1.0$~${\rm cm^{-3}}$ (thick red dashed and dotted lines). 
Furthermore,
 the events from rarefied medium, $n_0=0.01$~${\rm cm^{-3}}$, viewed off-axis ($\theta_v=0.2$~rad) can be also observed by 
CTA/4~LSTs
(thin blue dashed line in Fig.~\ref{CTA}(a)).
For the source redshift $z=0.5$,
CTA/4~LSTs
will detect the VHE gamma-rays from events with $n_0=1.0$~${\rm cm^{-3}}$ and $\theta_v=0.2$~rad (thick red dashed line in Fig.~\ref{CTA}(b)).
Due to the EBL attenuation, the 0.3~TeV flux from the source at $z=0.5$ becomes about an order of magnitude smaller than the unabsorbed one.
Orphan afterglows with higher redshifts ($z>0.5$) are hard to be detected with 
CTA/4~LSTs.

We roughly estimate the 
detection rate of orphan afterglows arising within $z=0.5$
by CTA/4~LSTs.
CTA/LST has a field of view of 
$2.5^\circ$ \protect\footnotemark[3]
with a duty cycle of 10\% \citep{CTA2019}.
We set the comoving rate density $R_{\rm GRB}\sim~300~{\rm Gpc}^{-3}~{\rm yr}^{-1}$ and the solid angle subtended by the direction to which the source is observed $f_\Omega=5.0\times10^{-3}$~sr.
Then, the expected detection rate by CTA within $z=0.5$ (luminosity distance $d_L\sim2.8$~Gpc) is simply given by
\begin{eqnarray}
\dot{N}_{\rm GRB}&\sim& \frac{4\pi}{3}~R_{\rm GRB}~d_L^3~f_\Omega ~\frac{(\theta_{\rm FOV})^2}{2}~DC \nonumber \\
&\sim& 0.15~{\rm yr^{-1}}~\left(\frac{R_{\rm GRB}}{300~{\rm Gpc^{-3}~yr^{-1}}}\right)\left(\frac{d_L}{2.8~{\rm Gpc}}\right)^3\left(\frac{f_\Omega}{5.0\times10^{-3}~{\rm sr}}\right)\left(\frac{\theta_{\rm FOV}}{2.5^\circ}\right)^2\left(\frac{DC}{10\%}\right)~~,
\label{eq:event}
\end{eqnarray}
where $\theta_{\rm FOV}$ is a field of view and $DC$ is the duty cycle.

We also calculate the X-ray, optical and radio  (16 and 343 GHz) fluxes with the same parameters as shown in Fig.~\ref{CTA}.
It is found from  Figs.~\ref{eROSITA}, \ref{LSST}, \ref{ALMA} and \ref{ngVLA} that
they are bright enough to be detected by eROSITA, Rubin LSST, ZTF, ALMA, ngVLA and SKA when $z = 0.5$.
Therefore, when 
CTA/4~LSTs detect
orphan afterglows, multi-wavelength observations are capable with
eROSITA, Rubin LSST, ZTF, ALMA, ngVLA and SKA.
\footnotetext[3]{https://zenodo.org/record/5499840\#.Y0TaWi0RodV}

%%%%%%%%%%%%%%%   Discussion   %%%%%%%%%%%%%%%
%%%%%%%%%%%%%%%%%%%%%%%%%%%%%%%%%%%
\section{Discussion}

We 
found 
that the observed VHE gamma-ray emission of GRB 190829A could be explained only by the SSC radiation.
If the isotropic-equivalent gamma-ray energy $E_{\rm iso,\gamma}(\theta_v)$ and peak energy $E_p(\theta_v)$ viewed on-axis ($\theta_v=0.0$~rad) are small, then, the total intrinsic energy is too small to explain the observed VHE gamma-ray flux of GRB 190829A by the SSC radiation \citep{BTZhang2021}.
However, we 
have supposed
that the jetted prompt emission of GRB 190829A would have been 
$E_{\rm iso,\gamma}(\theta_v=0.0$~rad) and $E_p(\theta_v=0.0$~rad) of typical long GRBs
if our narrow jet were
viewed on-axis.
Since the total intrinsic energy had large, the VHE gamma-ray light curve was explained only by the SSC emission (magenta solid line in Fig.~\ref{lightcurve}).

As seen in Fig.~\ref{lightcurve}, our numerical radio (99.8 GHz) and VHE gamma-ray fluxes of GRB 190829A were sometimes dimmer than the observational result. 
However, they are only within a factor of two, and this difference may come from 
the limitation of the simple afterglow model. 
For example, if microphysics parameters $\epsilon_e$, $\epsilon_B$ and $f_e$ have time-dependent, this problem may be solved. 
A small density fluctuation may reconcile them.
Our theoretical radio flux 
at
1.3 GHz overshot the observed 
one
between $2.0\times10^{5}$~s $< t <$ $6.0\times10^{5}$~s.
The excess 
was
also within a factor of two and it may solve in the same manner.
In \citet{Sato2021}, we calculated the SSC cooling in the Thomson limit and fixed as $f_e=1.0$.
However, in this paper, the Klein-Nishina effect and the case of $f_e\neq1.0$ were considered.
The late ($t\sim10^{6-7}$~s) X-ray and radio (1.3 GHz and 15.5 GHz) emissions of GRB 190829A have been improved (red, orange and green lines in Fig.~\ref{lightcurve}).

\begin{figure}
\centering
\begin{minipage}{0.3\linewidth}
\centering
\includegraphics[height=0.75\textwidth]{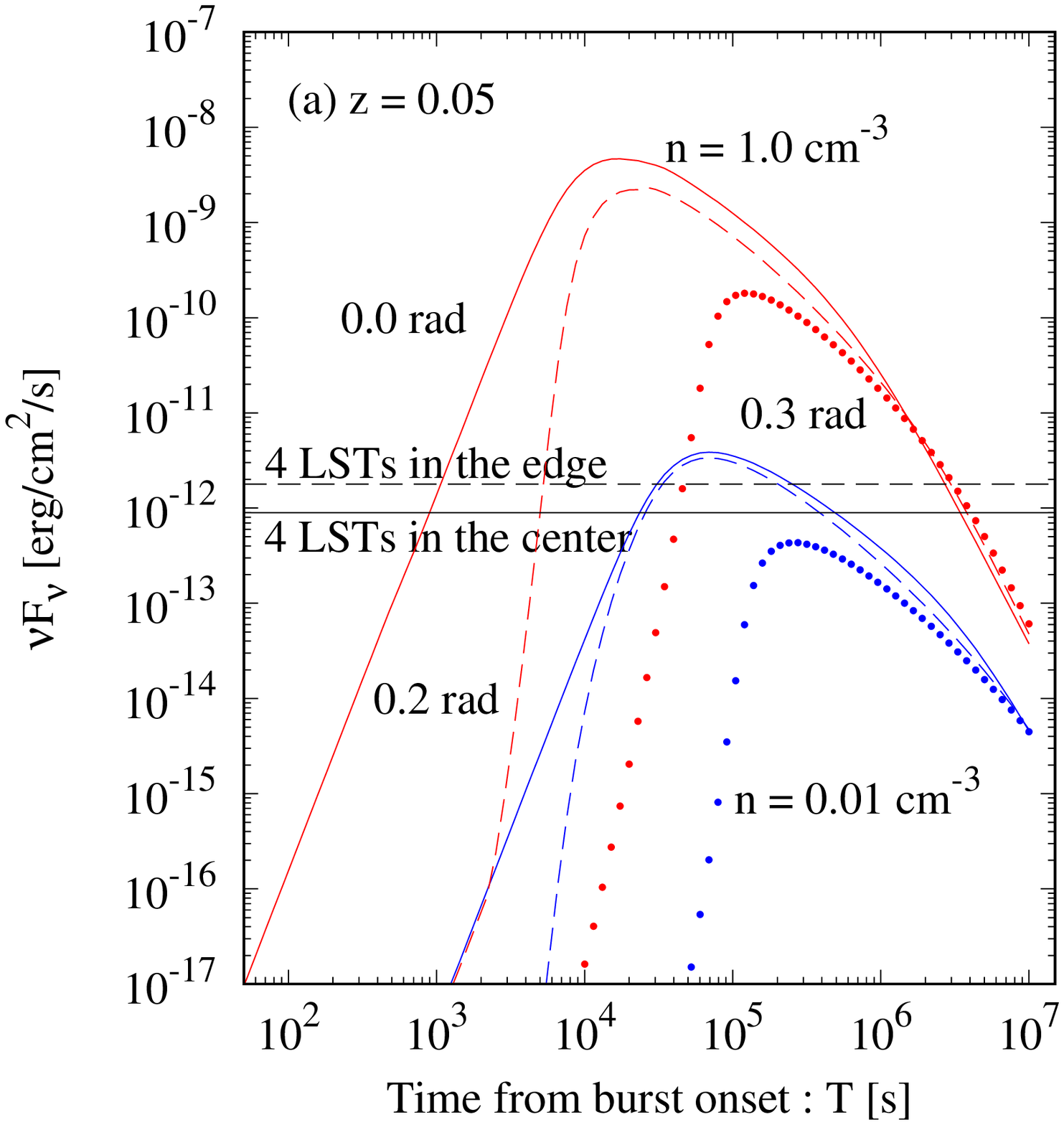}
\end{minipage}
\begin{minipage}{0.3\linewidth}
\centering
\includegraphics[height=0.75\textwidth]{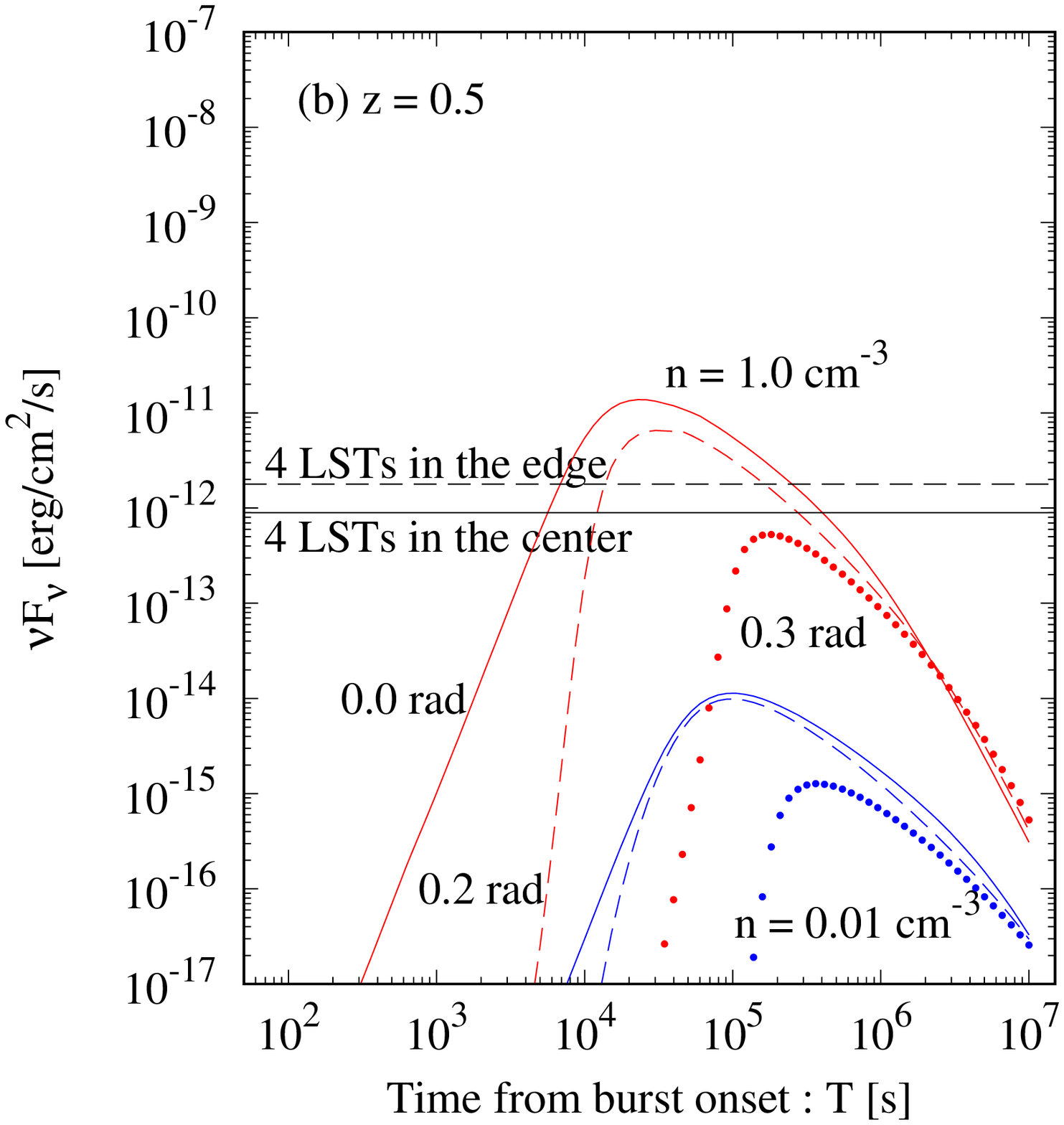}
\end{minipage}
\caption{
VHE gamma-ray light curves at $h\nu$ =0.3~TeV from the 'wide jet' ($\theta_0$ =  $0.1$~rad, $E_{\rm{iso, K}}$ =  $1.0 \times 10^{53}$~erg, $\Gamma_0$ = $20$,~$p$ = $2.2$, $\epsilon_B$ = $1.0 \times 10^{-5}$, $\epsilon_e$ = $0.29$, and $f_e$ =  $0.35$) at $z$ = $0.05$~(panel (a)) and $0.5$~(panel (b)).
The thick red and thin blue lines show the results for the ISM density $n_0$ = $1.0$ and $0.01$~${\rm cm^{-3}}$, respectively.
The solid, dashed and dotted lines are for cases of viewing angle $\theta_v$ = $0.0$, $0.2$ and $0.3$~rad, respectively.
Sensitivities of CTA/4~LSTs at the center (black solid line) and CTA/4~LSTs at the near edge (black dashed line) of the observation field of view are shown assuming an exposure time of three hours
\protect\footnotemark[3].
}
\label{CTA}
\end{figure}
\begin{figure}
\centering
\begin{minipage}{0.3\linewidth}
\centering
\includegraphics[height=0.75\textwidth]{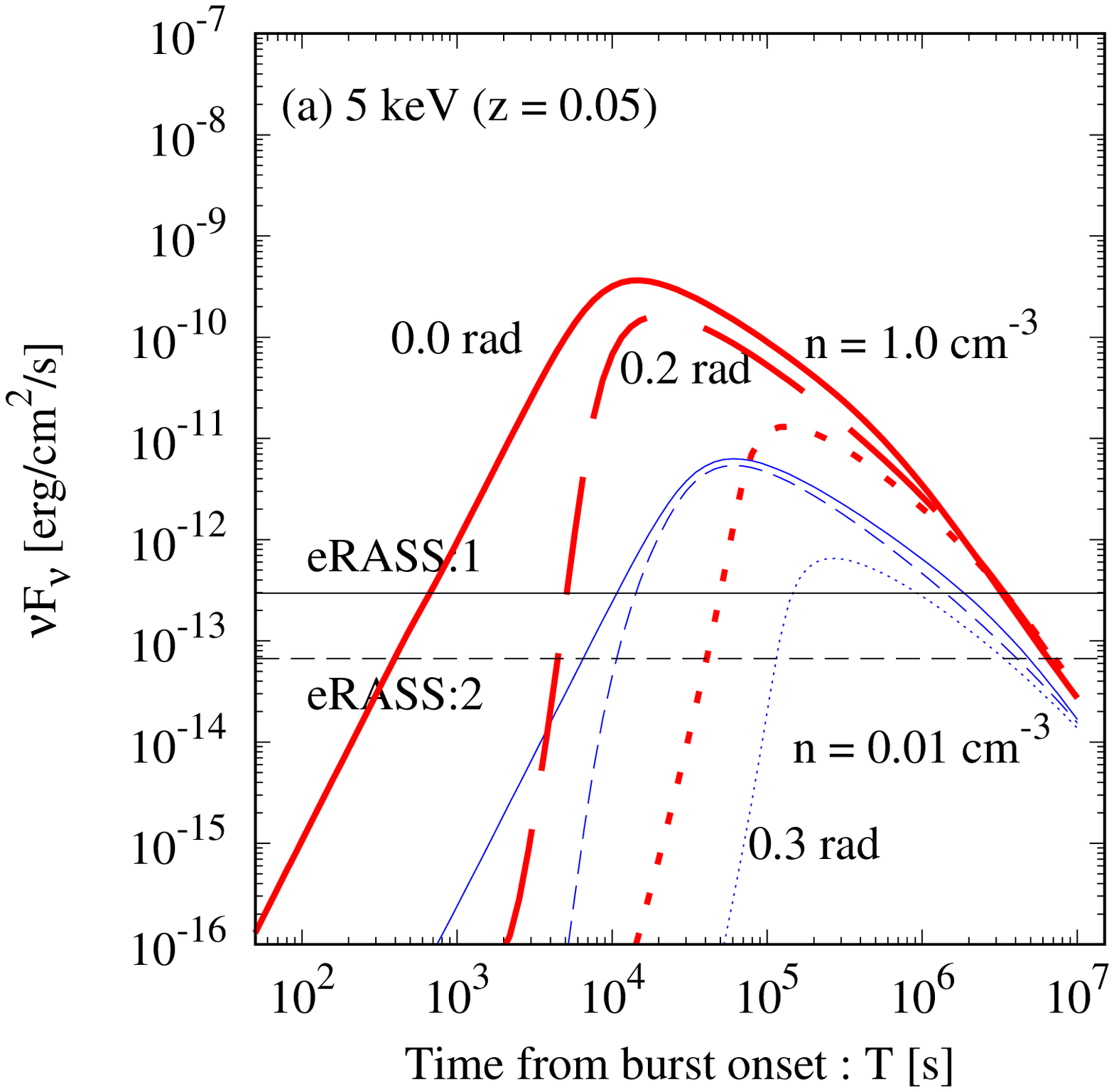}
\end{minipage}
\begin{minipage}{0.3\linewidth}
\centering
\includegraphics[height=0.75\textwidth]{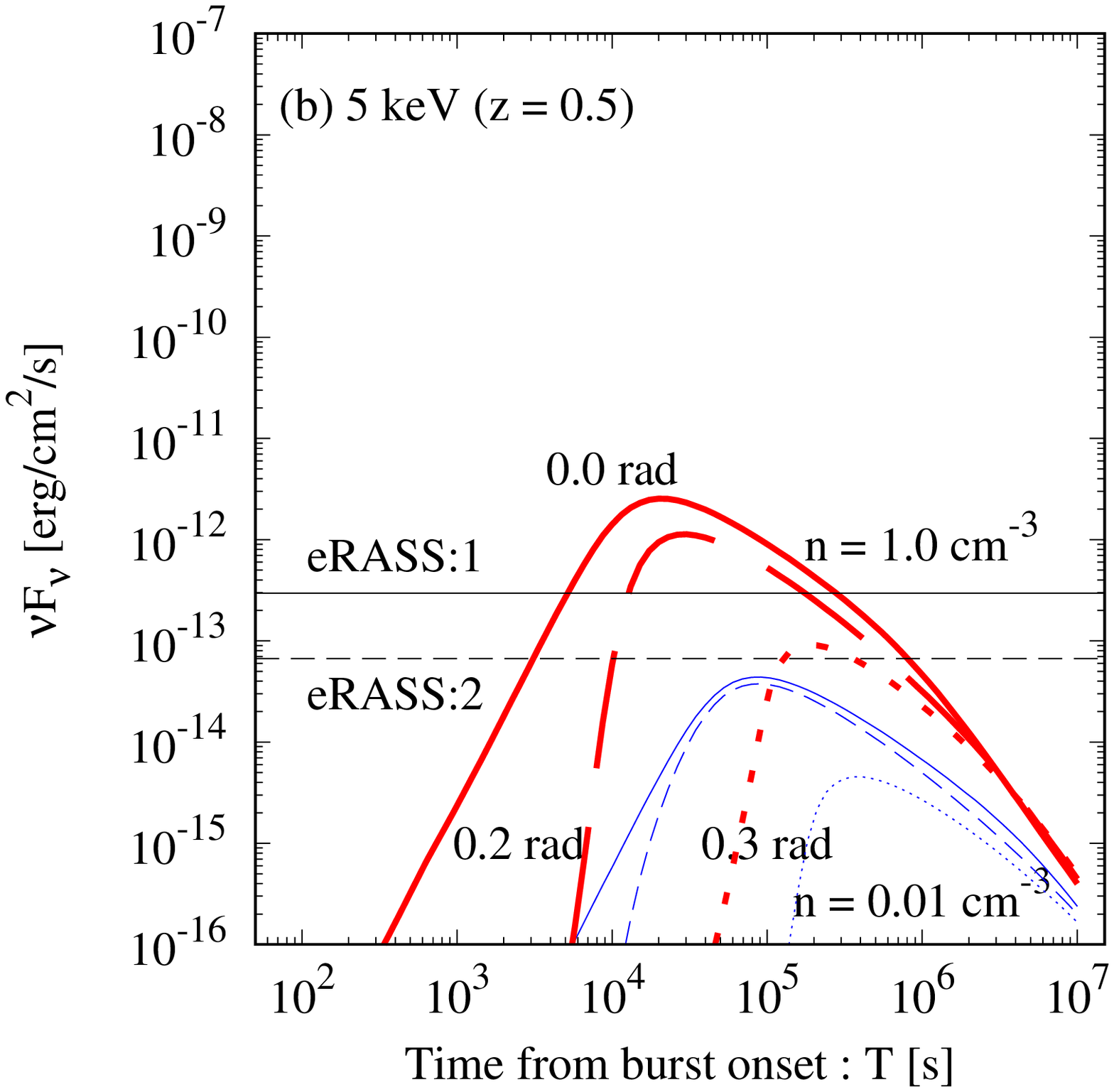}
\end{minipage}
\caption{
X-ray light curves at $h\nu$ =5~keV from the 'wide jet' ($\theta_0$ =  $0.1$~rad, $E_{\rm{iso, K}}$ =  $1.0 \times 10^{53}$~erg, $\Gamma_0$ = $20$, $p$ = $2.2$, $\epsilon_B$ = $1.0 \times 10^{-5}$, $\epsilon_e$ = $0.29$, and $f_e$ = $0.35$) at $z$ = $0.05$~(panel (a)) and $0.5$~(panel (b)).
The thick red and thin blue lines show the results for the ISM density $n_0$ = $1.0$ and $0.01$~${\rm cm^{-3}}$, respectively.
The solid, dashed and dotted lines are for cases of viewing angle $\theta_v$ = $0.0$, $0.2$ and $0.3$~rad, respectively.
Sensitivities of eRASS:~1 (black solid line) and eRASS:~2 (black dashed line) are shown assuming an exposure time of $500$~s \citep{Merloni2012}.
}
\label{eROSITA}
\end{figure}
\begin{figure}
\centering
\begin{minipage}{0.3\linewidth}
\centering
\includegraphics[height=0.75\textwidth]{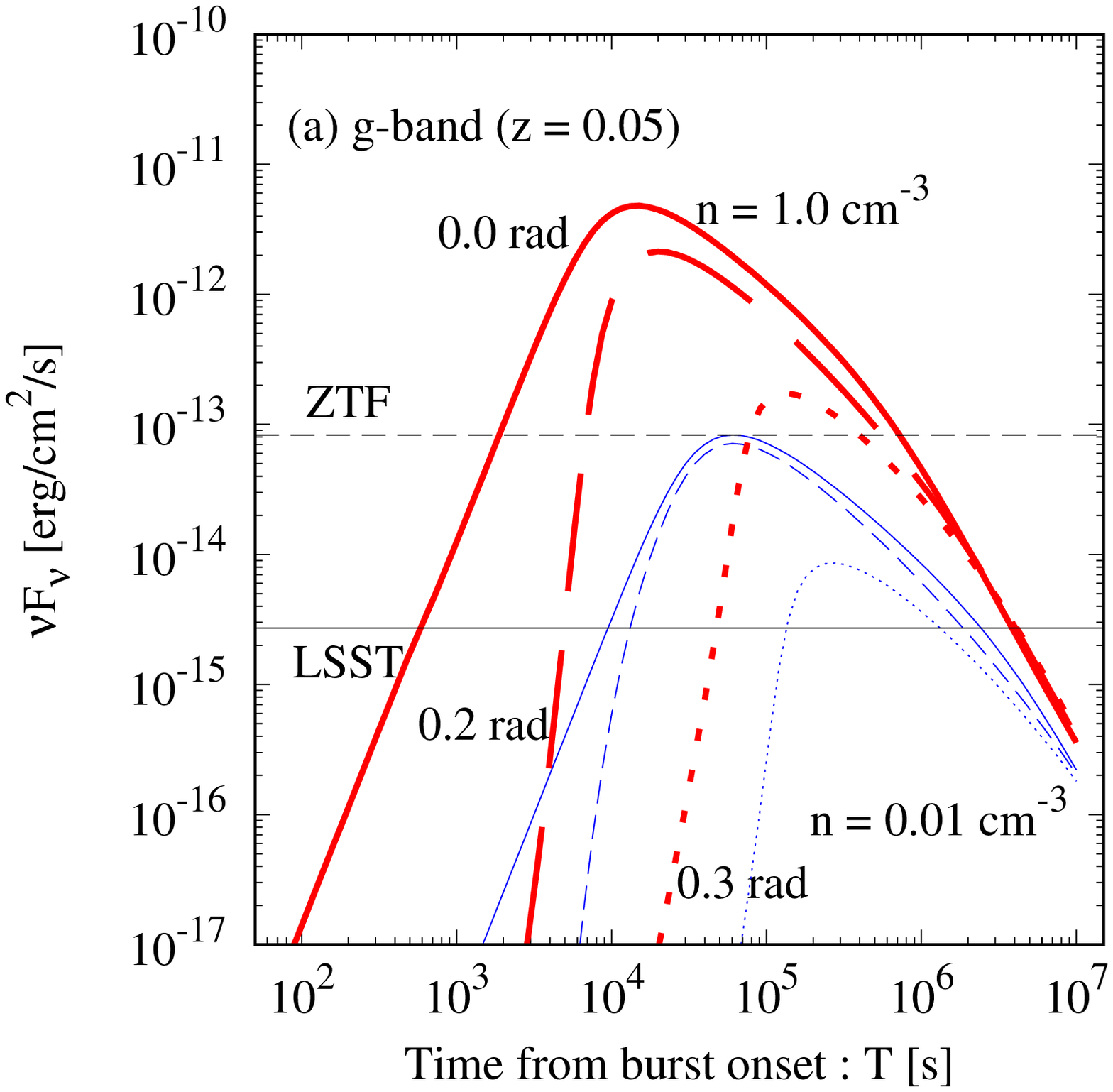}
\end{minipage}
\begin{minipage}{0.3\linewidth}
\centering
\includegraphics[height=0.75\textwidth]{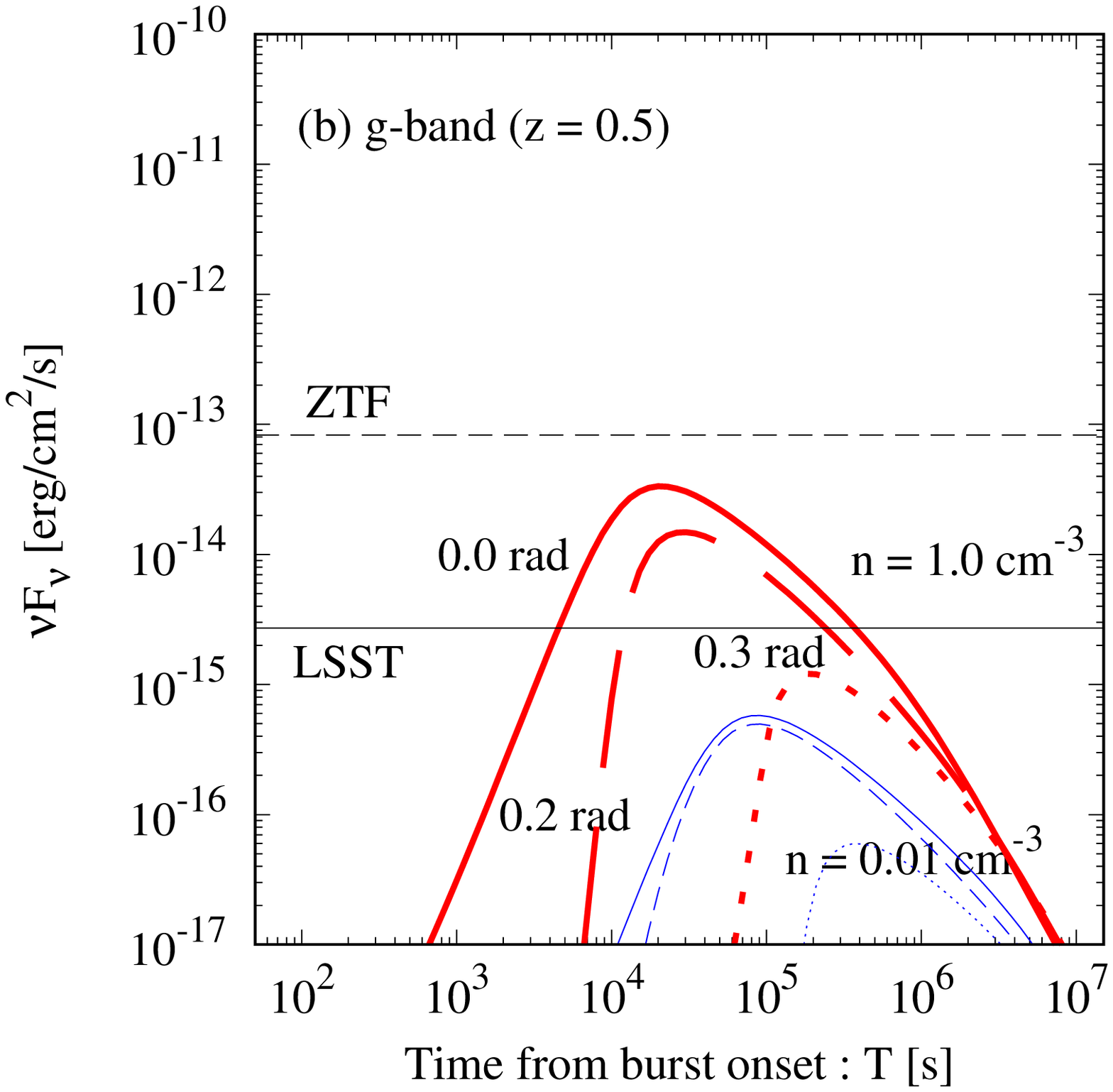}
\end{minipage}
\caption{
Optical light curves at g-band from the 'wide jet' ($\theta_0$ =  $0.1$~rad, $E_{\rm{iso, K}}$ =  $1.0 \times 10^{53}$~erg, $\Gamma_0$ = $20$, $p$ = $2.2$, $\epsilon_B$ = $1.0 \times 10^{-5}$, $\epsilon_e$ = $0.29$, and $f_e$ = $0.35$) at $z$ = $0.05$~(panel (a)) and $0.5$~(panel (b)).
The thick red and thin blue lines show the results for the ISM density $n_0$ = $1.0$ and $0.01$~${\rm cm^{-3}}$, respectively.
The solid, dashed and dotted lines are for cases of viewing angle $\theta_v$ = $0.0$, $0.2$ and $0.3$~rad, respectively.
Sensitivities of Rubin LSST (black solid line) \citep{Ivezic2019} and ZTF (black dashed line) \citep{Bellm2019} are shown assuming an exposure time of $140$~s.
}
\label{LSST}
\end{figure}
\begin{figure}
\centering
\begin{minipage}{0.3\linewidth}
\centering
\includegraphics[height=0.75\textwidth]{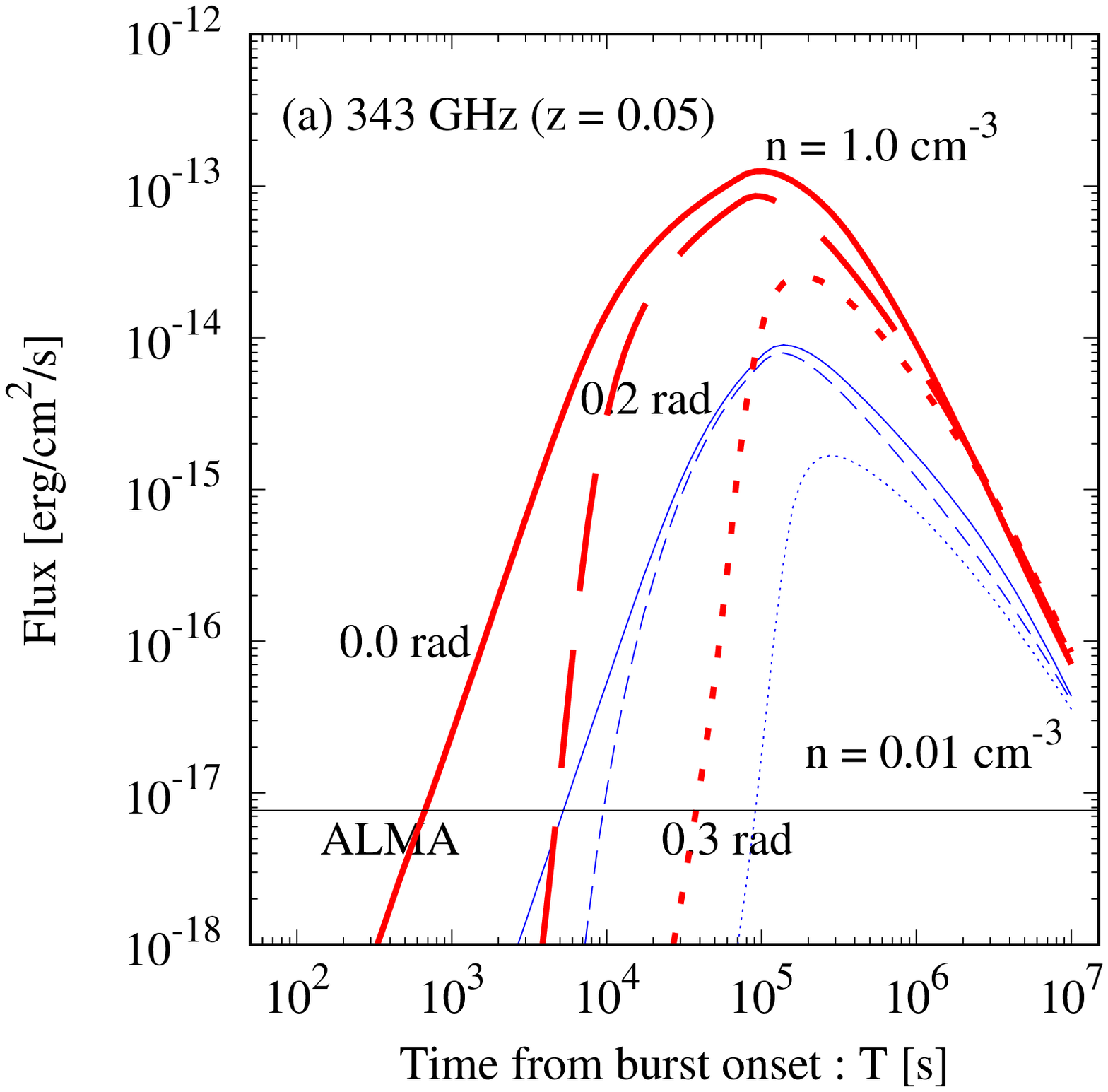}
\end{minipage}
\begin{minipage}{0.3\linewidth}
\centering
\includegraphics[height=0.75\textwidth]{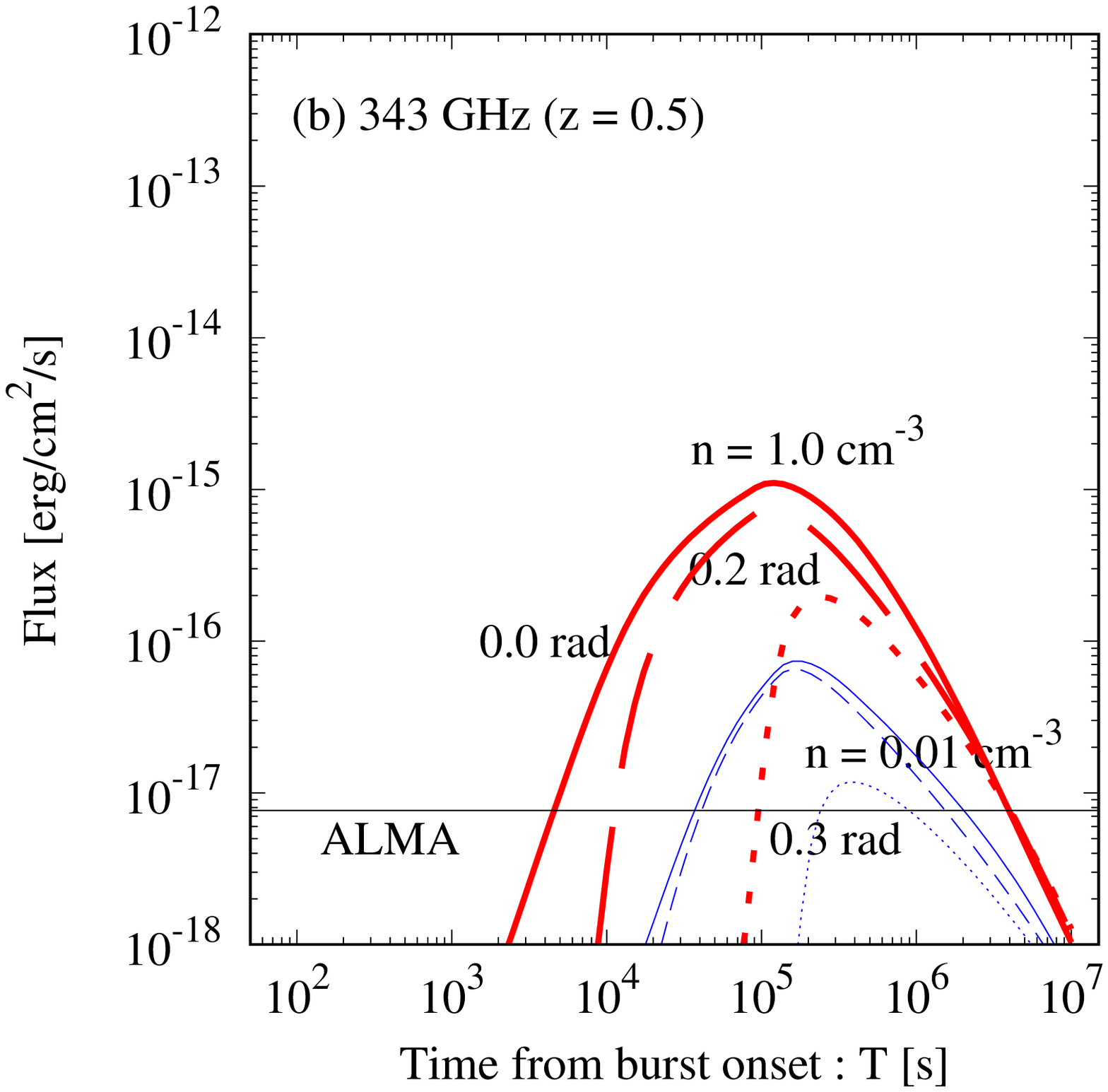}
\end{minipage}
\caption{
Radio light curves at 343 GHz from the 'wide jet' ($\theta_0$ =  $0.1$~rad, $E_{\rm{iso, K}}$ =  $1.0 \times 10^{53}$~erg, $\Gamma_0$ = $20$, $p$ = $2.2$, $\epsilon_B$ = $1.0 \times 10^{-5}$, $\epsilon_e$ = $0.29$, and $f_e$ = $0.35$) at $z$ = $0.05$~(panel (a)) and $0.5$~(panel (b)).
The thick red and thin blue lines show the results for the ISM density $n_0$ = $1.0$ and $0.01$~${\rm cm^{-3}}$, respectively.
The solid, dashed and dotted lines are for cases of viewing angle $\theta_v$ = $0.0$, $0.2$ and $0.3$~rad, respectively.
Sensitivity of ALMA (black solid line) 
is shown assuming an exposure time of $100$ hours \protect\footnotemark[4].
}
\label{ALMA}
\end{figure}
\footnotetext[4]{https://asa.alma.cl/SensitivityCalculator/}
\begin{figure}
\centering
\begin{minipage}{0.3\linewidth}
\centering
\includegraphics[height=0.75\textwidth]{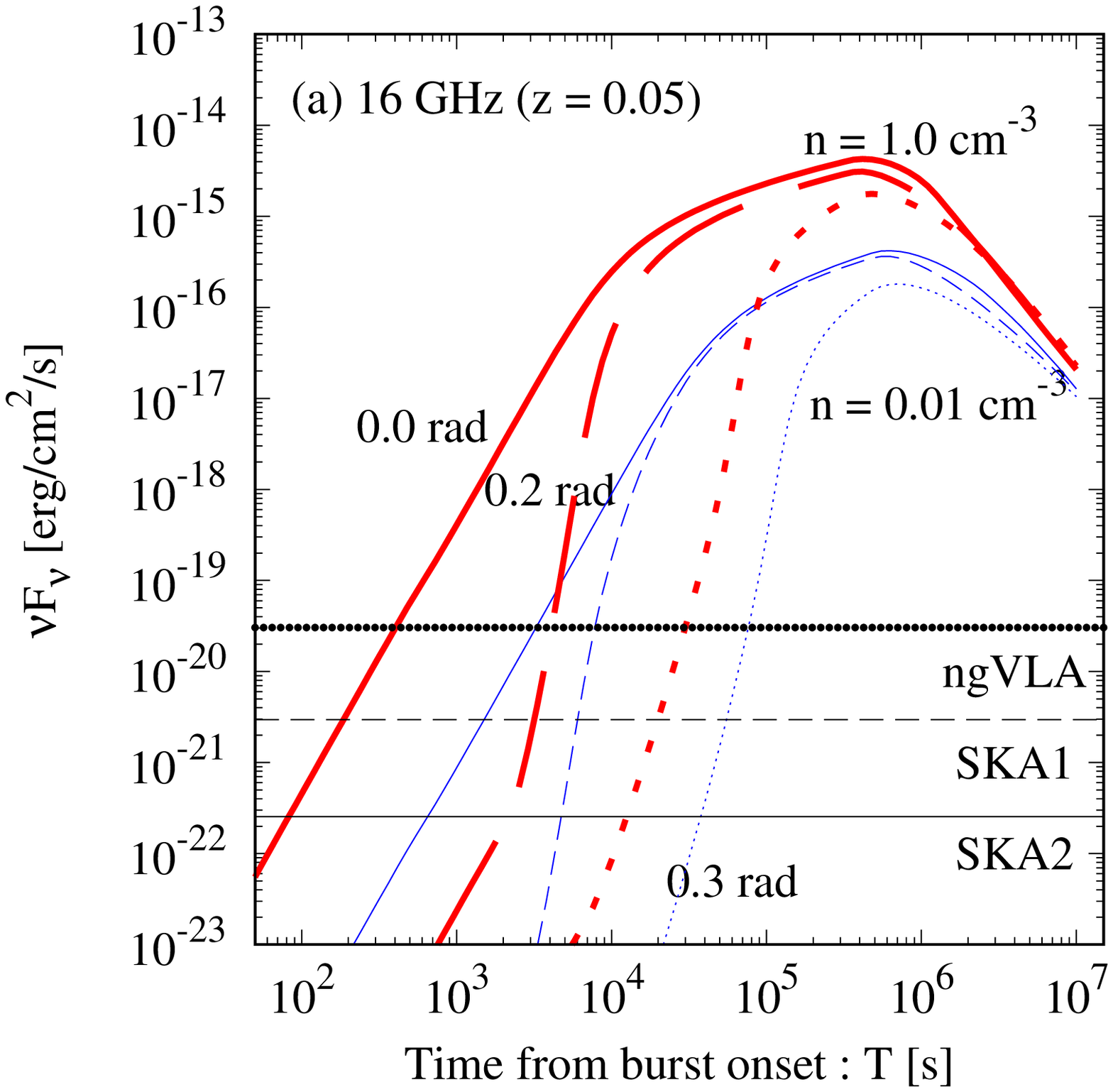}
\end{minipage}
\begin{minipage}{0.3\linewidth}
\centering
\includegraphics[height=0.75\textwidth]{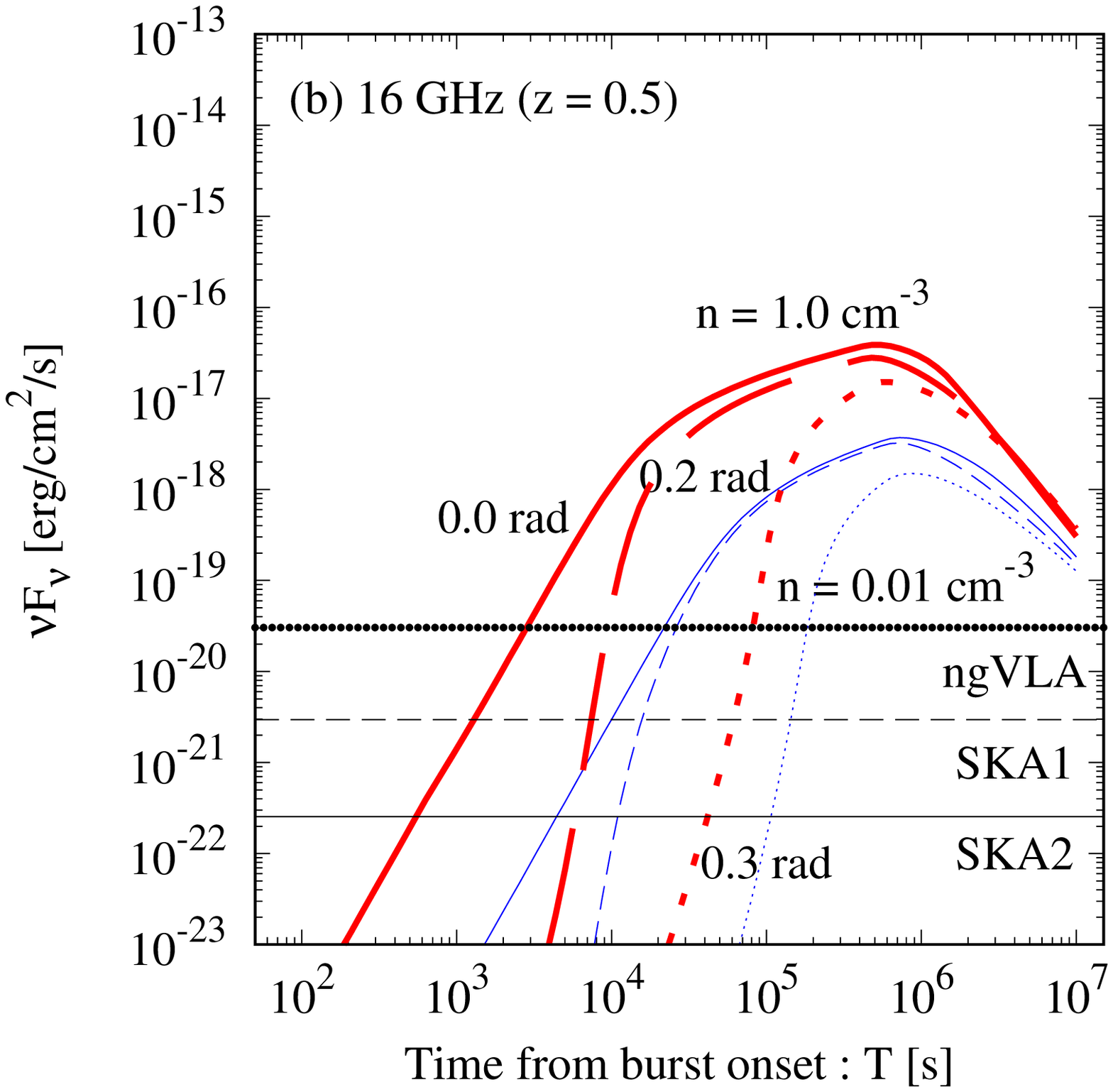}
\end{minipage}
\caption{
Radio light curves at 16 GHz from the 'wide jet' ($\theta_0$ =  $0.1$~rad, $E_{\rm{iso, K}}$ =  $1.0 \times 10^{53}$~erg, $\Gamma_0$ = $20$, $p$ = $2.2$, $\epsilon_B$ = $1.0 \times 10^{-5}$, $\epsilon_e$ = $0.29$, and $f_e$ = $0.35$) at $z$ = $0.05$~(panel (a)) and $0.5$~(panel (b)).
The thick red and thin blue lines show the results for the ISM density $n_0$ = $1.0$ and $0.01$~${\rm cm^{-3}}$, respectively.
The solid, dashed and dotted lines are for cases of viewing angle $\theta_v$ = $0.0$, $0.2$ and $0.3$~rad, respectively.
Sensitivities of ngVLA (black dotted line) \citep{Selina2018}, SKA1 (black dashed line) and SKA2 (black solid line) \citep{Chen2021} are shown assuming an exposure time of $100$ hours.
}
\label{ngVLA}
\end{figure}

After $5.0\times10^6$~s, the observed flux in 15.5 GHz of GRB 190829A was brighter than our result (see Fig.~\ref{lightcurve}), and it may be a radio supernova.
Type-Ic Supernova SN 2019oyw was associated with GRB 190829A \citep{Hu2021}.
The peak luminosity of 
a
typical type-Ic supernova is about $10^{28-29}~{\rm erg~s^{-1}~Hz^{-1}}$ at $t\sim10^7$~s in radio band \citep{Bietenholz2021}
At $5.0\times10^6$~s, the luminosity of GRB 190829A in 15.5 GHz band is about $10^{29}~{\rm erg~s^{-1}}$, and it is consistent with the typical peak luminosity of 
the
radio supernovae.

VLBI observations of GRB 190829A gave a constraint, $E_{\rm iso}/n_0 < 10^{56}$~${\rm erg~cm^3}$ \citep{Salafia2021}.
For our narrow and wide jet parameters, we get $E_{\rm iso}/n_0 = 4.0\times10^{55}$~${\rm erg~cm^3}$ and $1.0\times10^{55}$~${\rm erg~cm^3}$, respectively, and they are consistent with the observational 
upper limit.
Furthermore, source size upper limits (FWHM) in radio bands were obtained \citep{Salafia2021}.
Here, we also simply estimate the apparent jet size as the following.
The bulk Lorentz factor of the jet is initially so high that the jet size is calculated by $2R/\Gamma$.
As the jet decelerates, the beaming effect becomes weak.
The jet size is estimated as $2R\theta_j$, where $\theta_j$ is the jet opening half-angle.
The jet full angular size is calculated as $2~{\min} \{ R/ \Gamma,~R\theta_j\}/d_A$, where $d_A$ is the angular diameter distance to the source.
The value of $2~{\min} \{ R/ \Gamma,~R\theta_j\}/d_A$, multiplied by 0.65, is roughly equal to the FWHM of the observed radio angular size for the uniform jet \citep{Salafia2021}.
It is found that our result and observed data are consistent within $2\sigma$ range.

The upper limit on the linear polarization in the optical band of GRB 190829A was detected \citep{Dichiara2022}.
In \citet{Dichiara2022}, they supposed that the value of the linear polarization in the off-axis viewing case 
was
larger than the 
observed
one.
They adopted the model of \citet{Rossi2004},
which assumed that 
the magnetic field generated by the Weibel instability is extremely anisotropic and turbulent.
However, the magnetic field structure in the shock front is unclear.
If the direction of the distribution of the magnetic field downstream of the shock is less anisotropic,
the polarization of their model may be lower by a factor of 3-4 
\citep{Shimoda2021,Kuwata2022}.
The downstream magnetic field may not be formed by the Weibel instability.
Then, the value of the polarization is much uncertain.

For GRB 190829A,
the Fermi-LAT obtained upper limits in the HE range \citep{Fraija2021}.
In our calculation, the
numerical synchrotron radiation is brighter than 
the
SSC emission at 1~GeV, and 
the
numerical SSC flux dominates over the synchrotron component at 10 GeV in both the narrow and wide jets.
Our numerical emissions both 1 and 10 GeV bands are dimmer than the sensitivity of Fermi-LAT.

Our two-component jet model in the cases of 
the
on-axis viewing ($\theta_v=0.0$~rad) but with different values of  $n_0$, $p$, $\epsilon_e$, $\epsilon_B$ and $f_e$ well explained the observed light curves of GRBs 180720B, 190114C and 201216C (Figs.~\ref{VHE}(a), (b) and (c)).
For GRB 180720B, the first optical data point overshot our numerical result as well as \citet{Wang2019}.
It may be the contribution from the reverse-shock emission \citep{Fraija2019b}.
After about $t\sim80$~s, the HE gamma-ray flux of GRB 180720B decays steeply \citep{Ronchi2020}.
The narrow jet well 
explained
the observed HE gamma-ray data.

In GRB 190114C, the
observed
HE gamma-ray 
flux was
brighter than 
our  numerically calculated one.
It is claimed that 
the prompt and reverse-shock emissions may contribute to the HE gamma-ray flux
\citep{Wang2019, Asano2020}.
The early ($t\sim10^2-10^3$~s) bright optical flux of GRB 190114C was observed \citep{Jordana2020,Shrestha2022}.
It is difficult for the forward shock emission from our two-component jet model to explain the early optical observation.
\citet{Jordana2020} and \citet{Shrestha2022} supposed that the early optical radiation may be 
the reverse-shock
emission.
In \citet{Wang2019} and \citet{Asano2020}, the early ($t\lesssim10^2$~s) 
observed X-ray (5~keV)
afterglow exceeds their numerical results, while our 
X-ray (5~keV)
emission from our narrow jet is consistent with observed data (red solid line in Fig.~\ref{VHE}(b)).
The late ($t\gtrsim6.0\times10^3$~s) observed optical flux is brighter than the numerical result in \citet{Wang2019}.
Our wide jet is 
better
consistent with the observed optical light curve (blue solid line in Fig.~\ref{VHE}(b)) than previous theoretical works.
Before 20~s, 
the numerically calculated flux at 500~keV 
is dimmer than the observed data.
It may be the contribution of the prompt emission \citep{MAGIC2019a}.

For GRB 201216C, the observed VHE gamma-ray flux has not been published.
At 57~s when MAGIC detected the VHE gamma-ray 
photons,
our numerical calculation 
gave
the energy flux, $\nu F_{\nu}\sim2.0\times10^{-9}~{\rm erg~cm^{-2}~s^{-1}}$, at 0.1 TeV (magenta solid line in Fig.~\ref{VHE}(c)).
If our numerical VHE gamma-ray flux of GRB 201216C 
was
consistent with the observed one, we confirm the validity of our 
two-component
jet model.
In this paper, the observed radio data of GRB 201216C was explained by a two-component jet model as well as \citet{Rhodes2022}.

As seen in the yellow solid lines of 
Figs.~\ref{VHE}(a) and (b), the HE gamma-ray light curves have the plateaus.
The HE gamma-ray light curves with the plateau phase have been reported \citet{Ajello2019}.
According to our model, the two-component jet model is needed to explain complex multi-wavelength afterglow light curves of VHE 
gamma-ray
GRBs.
Furthermore, the values of $\epsilon_B\sim10^{-5}-10^{-4}$ and $E_{\rm iso,~K}\sim10^{53}$~erg are required for VHE gamma-ray emissions.
The VHE gamma-ray events may have unique features.

We have supposed the two-component jet,
which had
 the narrow-fast ($\theta_0 = 0.015$~rad and $\Gamma_0=350$) and 
wide-slow
($\theta_0 = 0.1$~rad and $\Gamma_0=20$) jets.
The two-component jet
model
has two different photon fields, so that 
inprinciple the external inverse-Compton (EIC) emission might be non-negligible.
If the two radiation regions have a large velocity difference,
then, the EIC mechanism may have an important role
\citep{Ghisellini2005}.
At the time when GRB 190829A was detected by H.E.S.S.,
however,
there was 
only a small
difference 
of
the 
bulk Lorentz factor
between our two jets \citep[see Fig. 3 in][]{Sato2021}.
The VHE gamma-ray fluxes from our two-component jet are less influenced by the EIC component, so that the observed data of GRB 190829A could be explained only by the SSC emission.
In our modeling, GRBs 180720B, 190114C and 201216C had the same $\theta_0$, $\Gamma_0$ and $E_{\rm iso,~K}$ as GRB 190829A, and $n_0$ of the three GRBs had similar
values.
Hence, the behaviors of the four-velocity $\Gamma\beta$ and radius $R$ of the jets in these events are similar to those of GRB 190829A.
The detection time of VHE gamma-ray photons of GRB 180720B ($t\sim3.6\times10^4$~s) was later than that of GRB 190829A ($t\sim2.0\times10^4$~s), so that the EIC component
less affects the VHE gamma-ray flux of GRB 180720B as well as GRB 190829A.
For GRB 190114C, from 57 s to 1591 s after the burst onset, the VHE gamma-ray emission was detected by MAGIC.
In this paper, we considered that GRB 190114C was viewed on-axis.
The bulk Lorentz factor of the narrow jet had still large from 57~s to 1591~s, so that the wide-jet synchrotron emission from behind is dim for the narrow jet due to the relativistic beaming effect.
At the narrow jet, the energy density synchrotron photon density from the wide jet is much smaller than one of the narrow jet by itself.
The EIC mechanism less affects the observational flux of GRB 190114C.

We confirmed that our two-component jet model was consistent with the observational results of GRBs 180720B, 190114C, 190829A and 201216C in \S~3.1 and 3.2.
The electron spectral index $p$ of the narrow and wide jets might be different 
from
each other, and then
we may observe the change of the spectral index from one to another.
It may be an indication of the existence of the two components. 
Another possibility is that we may observe the dip and/or bumps in light curves, which are difficult to be explained by the single component jet
\citep{Beniamini2020,Duque2022}.
Furthermore, late radio observations may be important to discriminate the two-component jet.
The narrow jet emissions decay rapidly, and they are followed by late-time wide jet emissions.

As shown in Tables~1, 2 and 3, the narrow and wide jets have different microphysics parameters, $\epsilon_e$, $\epsilon_B$ and $f_e$.
They might depend on the properties of circumburst matter like the number density.
However, at present, it is still unknown how their values are determined, and therefore, the dependence on the ambient medium is not clarified.
There are several possibilities for making an anisotropic medium into which the two jets penetrate, resulting in their different microphysics parameters.
If a progenitor star is rotating, a unique environment might be built along the rotational axis.
Then, the magnetic fields and density could depend on the angle from the rotational axis. 
In addition, differences in the deceleration of the narrow and wide jets may affect microphysics parameters in the forward-shocked region because the non-linear evolution of the Rayleigh-Taylor instability affects the forward-shocked region. 
Inherent ejecta properties like magnetization parameters could be transferred into the emitting thin shell by mixing the ejecta and circumburst material.
If the ejecta of the narrow and wide jets is different, the value of microphysics parameters of the narrow and wide jets could have different values.

In \S~3.3, we calculated the event rate of the orphan afterglow emission by 
CTA/4~LSTs
during the observations of other targets.
Orphan afterglows are more likely to be detected at the edge of the field of view of 
CTA/4~LSTs.
The sensitivity in the edge of 
the
field of view of 
CTA/4~LSTs
is about two times lower than that of the center.
Since the VHE
gamma-ray
flux exceeded the sensitivity
in the edge
of 
CTA/4~LSTs
(thick red and thin blue lines Figs.~\ref{CTA}(a) and (b)), the event rate may not be largely affected. 

The light curves viewed on-axis ($\theta_v=0.0$~rad) and off-axis ($\theta_v=0.2$ and $0.3$~rad) in 
the
VHE gamma-ray, X-ray and optical bands have achromatic peaks (see solid, dashed and dotted lines in Figs.~\ref{CTA}, \ref{eROSITA} and \ref{LSST}).
In the on-axis viewing case, the light curves have achromatic peaks when the transitions from 
the
free-expansion to the Blandford-McKee phase \citep{Sari1997}.
The observer time of the flux maximum is analytically given by \citet{Sari1997}
\begin{eqnarray}
t_{\rm peak, on} &\sim& (1+z)\left(\frac{3E_{\rm iso, K}}{32\pi n_0m_pc^5\Gamma_0^8}\right)^{1/3} \nonumber \\
&\sim& 1.5\times10^4~\left(\frac{1+z}{1.05}\right)\left(\frac{E_{\rm iso, K}}{10^{53}~{\rm erg}}\right)^{1/3}\left(\frac{n_0}{1.0~{\rm cm^{-3}}}\right)^{-1/3}\left(\frac{\Gamma_0}{20}\right)^{-8/3}~{\rm s}~~.
\label{eq:onpeak}
\end{eqnarray}
The off-axis afterglows show the rising part because of the relativistic beaming effect.
The observer time of the peak in the light curve for $\theta_v=0.2$~rad is estimated by 
\begin{eqnarray}
t_{\rm peak, off} &\sim&(1+z)\left(\frac{3E_{{\rm iso,K}}}{32\pi n_0m_pc^5}\right)^{1/3}\left(\frac{\theta_v-\theta_0}{2}\right)^{8/3} \nonumber \\
&\sim&  3.0\times10^4~\left(\frac{1+z}{1.05}\right)\left(\frac{E_{\rm iso, K}}{10^{53}~{\rm erg}}\right)^{1/3}\left(\frac{n_0}{1.0~{\rm cm^{-3}}}\right)^{-1/3}\left(\frac{\theta_v-\theta_0}{0.1~{\rm rad}}\right)^{8/3}~{\rm s}~~.
\label{eq:offpeak}
\end{eqnarray}
In the case of $\theta_v=0.3$~rad, 
we obtain
$t_{\rm peak, off} \sim 1.9\times10^5$~s.
The emissions viewed
with
$\theta_v=0.0$ and $0.2$~rad 
have
maximum almost at the same time,
while the observer time at the peak in the case of $\theta_v=0.3$~rad 
is
later than the on-axis viewing ($\theta_v=0.0$~rad) case.
Other combinations of $n_0$ and $z$ 
behave
in the same manner.
In the radio bands (16 and 343~GHz), each light curve has a peak when the typical frequency $\nu_m$ crosses 16 and 343~GHz 
bands,
respectively.
In particular, for $z=0.05$ and $n_0$~=~1.0~${\rm cm^{-3}}$, 
$\nu_m$
intersects 16~GHz at $t\sim5.0\times10^5$~s, and 
then, our numerical result in 16~GHz takes maximum
(thick red lines in Figs.~\ref{ALMA}(a) and \ref{ngVLA}(a)).

\begin{figure*}
\centering 
\includegraphics[width=0.3\textwidth]{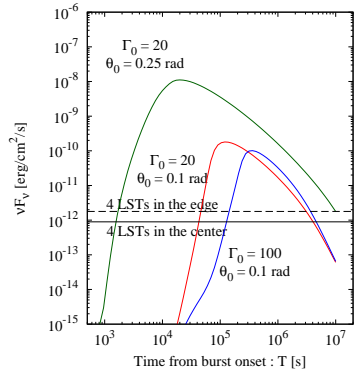}
\caption{
VHE gamma-ray light curves at $h\nu =0.3$~TeV.
The value of $\theta_0$ and $\Gamma_0$ is changed (thick red solid line: $\theta_0=0.1$~rad and $\Gamma_0=20$, thin blue dashed line: $\theta_0=0.1$~rad and $\Gamma_0=100$, thin green solid line: $\theta_0=0.25$~rad and $\Gamma_0=20$).
We fix $z=0.05$, $n_0$ =  1.0~${\rm cm^{-3}}$, $\theta_v$ =  0.3~rad, $E_{\rm{iso, K}}$ =  $1.0 \times 10^{53}$ erg, $\epsilon_B$ =  $1.0 \times 10^{-5}$, $\epsilon_e$ =  0.29, and $f_e$ =  0.35.
Sensitivities of 
CTA/4~LSTs at the center (black solid
line) and CTA/4~LSTs at the edge (black dashed line) of the field of view
are shown assuming an exposure time of three hours 
\protect\footnotemark[3].
}
\label{compara}
\end{figure*}

We considered the jets which had the same wide jet parameters ($p$, $E_{\rm iso}$, $\epsilon_e$, $\epsilon_B$ and $f_e$) but different initial jet opening half-angle ($\theta_0=0.1$ and 0.25 rad) and initial jet Lorentz factor ($\Gamma_0=100$ and 20) with $n_0$~=~1.0~${\rm cm^{-3}}$ within $z = 0.05$.
For our wide jet ($\Gamma_0=20$ and $\theta_0=0.1$~rad), the typical jet ($\Gamma_0=100$ and $\theta_0=0.1$~rad) and the jet with $\Gamma_0=20$ and $\theta_0=0.25$~rad, 
the peak of observed light curves at 0.3 TeV is at $t_{{\rm peak,off}} \sim 9.0\times10^4$~s, $3.0\times10^5$~s and $1.5\times10^4$~s, respectively (thick red solid, thin blue solid and thin green dashed lines in Fig.~\ref{compara}).
The peak times for the three jets are different.
If the light curve which has the observer time of the flux maximum at $t\sim10^5$~s (thick red solid line in Fig.~\ref{compara}) is detected by
CTA/4~LSTs
with $n_0$~=~1.0~${\rm cm^{-3}}$ within $z = 0.05$, we could confirm the existence of the wide jet.

%%%%%%%%%%%%%%%   Summary   %%%%%%%%%%%%%%%
%%%%%%%%%%%%%%%%%%%%%%%%%%%%%%%%%%%
\section{Summary}

In this paper, we have calculated afterglow emissions using the two-component jet model as given by \citet{Sato2021} to explain the new observational results of the VHE gamma-ray and radio bands (5.5 GHz and 99.8 GHz) of GRB 190829A.
In our previous work \citep{Sato2021}, the number fraction of accelerated electrons was fixed as $f_e = 1.0$, while in this paper, we chose
$f_e = 0.2$ and $0.35$
for the narrow and wide jets, respectively.
As a result, our numerical results in 1.3 and 15.5 GHz bands 
have been improved. 
In this paper, the Klein-Nishina effect in calculating the SSC process has been considered, so that our numerical X-ray light curve better matched than the results of \citet{Sato2021}.
The multi-wavelength afterglows of GRB 190829A were well explained by our two-component jet model (Fig.~\ref{lightcurve}).
Furthermore, the afterglow emissions from our two-component jet were consistent with the observed data of GRBs 180720B, 190114C and 201216C.
The two-component 
jet, with the narrow-fast ($\theta_0=0.015$~rad and $\Gamma_0=350$) and %
wide-slow
($\theta_0=0.1$~rad and $\Gamma_0=20$) jets, may be required to explain the multi-wavelength afterglows of the VHE gamma-ray events.
According to our two-component jet model, the detection rate of orphan afterglows within $z=0.5$ by CTA/4~LSTs was about $0.1~{\rm yr^{-1}}$.

\section*{Acknowledgements}
We thank 
Shotaro~Abe,
Kento~Aihara,
Katsuaki~Asano,
Giancarlo~Ghirlanda,
Gabriel~Ghisellini,
Kazuaki~Hashiyama,
Shotaro~Ide,
Kunihito~Ioka, 
Kyohei~Kawaguchi,
Shota~Kisaka,
Asuka~Kuwata,
Lara~Nava,
Koji~Noda,
Yoshihiro~Okutani,
Takayuki~Saito,
Takanori~Sakamoto,
Haruka~Sakemi,
Om~Sharan~Salafia,
Kensyo~Sei,
Kazuyoshi~Tanaka,
Kenta~Terauchi,
and 
Kenji~Toma
for valuable comments.
This research has made use of the CTA instrument response functions provided by the CTA Consortium and Observatory, see https://www.cta-observatory.org/science/cta-performance/ (version prod5 v0.1; \citet{CTA2021}) for more details.
We also thank the referee for his or her helpful comments to substantially improve the paper. 
This research
was partially supported by JSPS KAKENHI Grant Nos. 22J20105 (YS), 20H01901 (KM), 20H05852 (KM), 19H01893 (YO), 21H04487 (YO) and 22H01251 (RY).
SJT is supported by the Sumitomo Foundation for basic science research projects (210629), Research Foundation for Opto-Science and Technology, and Aoyama Gakuin University Research Institute.
The work of KM is supported by the NSF Grant No. AST-1908689, No. AST-2108466 and No. AST-2108467.
YO is supported by Leading Initiative for Excellent Young Researchers, MEXT, Japan.

%\bibliography{mybibfile}

\begin{thebibliography}{99}
\bibitem[\protect\citeauthoryear{Acciari et al.}{2021}]{Acciari2021} 
Acciari V.~A., et al., 2021, ApJ, 908, 90

\bibitem[\protect\citeauthoryear{Abdalla et al.}{2019}]{Abdalla2019} 
Abdalla H., et al. 2019, Natur, 575, 464

\bibitem[\protect\citeauthoryear{Ajello et al.}{2019}]{Ajello2019}
Ajello M., et al., 2019, ApJ, 878, 52


\bibitem[\protect\citeauthoryear{Asano, Murase, \& Toma}{2020}]{Asano2020}
Asano K., Murase K., Toma K., 2020, ApJ, 905, 105

\bibitem[\protect\citeauthoryear{Bellm et al.}{2019}]{Bellm2019} 
Bellm E.~C., et al., 2019, PASP, 131, 018002

\bibitem[\protect\citeauthoryear{Beniamini et al.}{2020}]{Beniamini2020} 
Beniamini P., Duque R., Daigne F., Mochkovitch R., 2020, MNRAS, 492, 2847

\bibitem[\protect\citeauthoryear{Bietenholz et al.}{2021}]{Bietenholz2021} 
Bietenholz M.~F., Bartel N., Argo M., Dua R., Ryder S., Soderberg A., 2021, ApJ, 908, 75

\bibitem[\protect\citeauthoryear{Blanch et al.}{2020a}]{Blanch2020a} 
Blanch O., et al., 2020, GCN, 28659, 1

\bibitem[\protect\citeauthoryear{Blanch et al.}{2020b}]{Blanch2020b} 
Blanch O., et al., 2020, ATel, 14275, 1

\bibitem[\protect\citeauthoryear{Blumenthal \& Gould}{1970}]{Blumenthal1970} 
Blumenthal G.~R., Gould R.~J., 1970, RvMP, 42, 237

\bibitem[\protect\citeauthoryear{Chand et al.}{2020}]{Chand2020} 
Chand V., et al. 2020, ApJ, 898, 42

\bibitem[\protect\citeauthoryear{Chen, Sming Tsai, \& Yuan}{2021}]{Chen2021} 
Chen Z., Sming Tsai Y.-L., Yuan Q., 2021, JCAP, 2021, 025

\bibitem[\protect\citeauthoryear{Cherenkov Telescope Array Consortium et al.}{2019}]{CTA2019} 
Cherenkov Telescope Array Consortium, et al., 2019, {\it Science with the Cherenkov Telescope Array}, World Scientific Publishing Co. Pte. Ltd., doi:10.1142/10986

\bibitem[\protect\citeauthoryear{Cherenkov Telescope Array Observatory and Cherenkov Telescope Array Consortium}{2021}]{CTA2021} 
Cherenkov Telescope Array Observatory and Cherenkov Telescope Array Consortium, 2021, {\it CTAO Instrument Response Functions - prod5 version v0.1}, Zenodo, doi:10.5281/zenodo.5499840


\bibitem[\protect\citeauthoryear{Derishev \& Piran}{2019}]{Derishev2019} 
Derishev E., Piran T., 2019, ApJL, 880, L27

\bibitem[\protect\citeauthoryear{Derishev \& Piran}{2021}]{Derishev2021} 
Derishev E., Piran T., 2021, ApJ, 923, 135

\bibitem[\protect\citeauthoryear{Dichiara et al.}{2022}]{Dichiara2022} 
Dichiara S., et al., 2022, MNRAS, 512, 2337

\bibitem[\protect\citeauthoryear{Donaghy}{2006}]{Donaghy2006} 
Donaghy T.~Q., 2006, ApJ, 645, 436

\bibitem[\protect\citeauthoryear{Duan \& Wang}{2019}]{Duan2019}
Duan M.-Y., Wang X.-G., 2019, ApJ, 884, 61

\bibitem[\protect\citeauthoryear{Duque et al.}{2022}]{Duque2022} 
Duque R., Beniamini P., Daigne F., Mochkovitch R., 2022, MNRAS, 513, 951


\bibitem[\protect\citeauthoryear{Eichler \& Waxman}{2005}]{Eichler2005} 
Eichler D., Waxman E., 2005, ApJ, 627, 861

\bibitem[\protect\citeauthoryear{Evans et al.}{2007}]{Evans2007} 
Evans P.~A., et al., 2007, A\&A, 469, 379

\bibitem[\protect\citeauthoryear{Evans et al.}{2009}]{Evans2009} 
Evans P.~A.,  et al., 2009, MNRAS, 397, 1177

\bibitem[\protect\citeauthoryear{Fraija et al.}{2019a}]{Fraija2019a} 
Fraija N., Barniol Duran R., Dichiara S., Beniamini P., 2019a, ApJ, 883, 162

\bibitem[\protect\citeauthoryear{Fraija et al.}{2019b}]{Fraija2019b} 
Fraija N., et al., 2019b, ApJ, 885, 29

\bibitem[\protect\citeauthoryear{Fraija et al.}{2019c}]{Fraija2019c} 
Fraija N., Dichiara S., Pedreira A.~C.~C. do E.~S., Galvan-Gamez A., Becerra R.~L., Barniol Duran R., Zhang B.~B., 2019c, ApJL, 879, L26

\bibitem[\protect\citeauthoryear{Fraija et al.}{2021}]{Fraija2021} 
Fraija N., Veres P., Beniamini P., Galvan-Gamez A., Metzger B.~D., Barniol Duran R., Becerra R.~L., 2021, ApJ, 918, 12

\bibitem[\protect\citeauthoryear{Franceschini, Rodighiero, \& Vaccari}{2008}]{Franceschini2008} 
Franceschini A., Rodighiero G., Vaccari M., 2008, A\&A, 487, 837

\bibitem[\protect\citeauthoryear{Gao et al.}{2013}]{Gao2013} 
Gao H., Lei W.-H., Zou Y.-C., Wu X.-F., Zhang B., 2013, NewAR, 57, 141

\bibitem[\protect\citeauthoryear{Gao et al.}{2022}]{Gao2022} 
Gao H.-X., et al., 2022, preprint, arXiv:2204.03823

\bibitem[\protect\citeauthoryear{Ghisellini, Tavecchio, \& Chiaberge}{2005}]{Ghisellini2005} 
Ghisellini G., Tavecchio F., Chiaberge M., 2005, A\&A, 432, 401

\bibitem[\protect\citeauthoryear{Gottlieb et al.}{2020}]{Gottlieb2020} 
Gottlieb O., Bromberg O., Singh C.~B., Nakar E., 2020, MNRAS, 498, 3320


\bibitem[\protect\citeauthoryear{Granot, Piran, \& Sari}{1999}]{Granot1999} 
Granot J., Piran T., Sari R., 1999, ApJ, 513, 679

\bibitem[\protect\citeauthoryear{Graziani, Lamb, \& Donaghy}{2006}]{Graziani2006} 
Graziani C., Lamb D.~Q., Donaghy T.~Q., 2006, AIPC, 836, 117

\bibitem[\protect\citeauthoryear{H.E.S.S. Collaboration et al.}{2021}]{HESS2021} 
H.E.S.S. Collaboration 2021, Sci, 372, 1081

\bibitem[\protect\citeauthoryear{Ho et al.}{2020}]{Ho2020} 
Ho A.~Y.~Q., et al., 2020, ApJ, 905, 98

\bibitem[\protect\citeauthoryear{Ho et al.}{2022}]{Ho2022} 
Ho A.~Y.~Q., et al., 2022, preprint (arXiv:2201.12366)

\bibitem[\protect\citeauthoryear{Hu et al.}{2021}]{Hu2021} 
Hu Y.-D., et al., 2021, A\&A, 646, A50

\bibitem[\protect\citeauthoryear{Huang et al.}{2000}]{Huang2000} 
Huang Y.~F., Gou L.~J., Dai Z.~G., Lu T. 2000, ApJ, 543, 90

\bibitem[\protect\citeauthoryear{Huang et al.}{2021}]{Huang2021} 
Huang X.-L., Wang Z.-R., Liu R.-Y., Wang X.-Y., Liang E.-W., 2021, ApJ, 908, 225

\bibitem[\protect\citeauthoryear{Huang et al.}{2020}]{Huang2020} 
Huang Y.-J., et al., 2020, ApJ, 897, 69

\bibitem[\protect\citeauthoryear{Huang}{2022}]{Huang2022} 
Huang Y., 2022, preprint (arXiv:2204.08208)

\bibitem[\protect\citeauthoryear{Ioka \& Nakamura}{2001}]{Ioka2001} 
Ioka K., Nakamura T., 2001, ApJL, 554, L163

\bibitem[\protect\citeauthoryear{Ivezi{\'c} et al.}{2019}]{Ivezic2019} 
Ivezi{\'c} {\v{Z}}., et al., 2019, ApJ, 873, 111

\bibitem[\protect\citeauthoryear{Jacovich, Beniamini, \& van der Horst}{2021}]{Jacovich2021} 
Jacovich T.~E., Beniamini P., van der Horst A.~J., 2021, MNRAS, 504, 528

\bibitem[\protect\citeauthoryear{Jordana-Mitjans et al.}{2020}]{Jordana2020} 
Jordana-Mitjans N., et al., 2020, ApJ, 892, 97

\bibitem[\protect\citeauthoryear{Kuwata et al.}{2022}]{Kuwata2022} 
Kuwata A., Toma K., Kimura S.~S., Tomita S., Shimoda J., 2022, preprint (arXiv:2208.09242)


\bibitem[\protect\citeauthoryear{MAGIC Collaboration et al.}{2019a}]{MAGIC2019a} 
MAGIC Collaboration 2019, Natur, 575, 455

\bibitem[\protect\citeauthoryear{MAGIC Collaboration et al.}{2019b}]{MAGIC2019b} 
MAGIC Collaboration, et al., 2019, Natur, 575, 459


\bibitem[\protect\citeauthoryear{Merloni et al.}{2012}]{Merloni2012} 
Merloni A., 2012, preprint (arXiv:1209.3114)

\bibitem[\protect\citeauthoryear{Meszaros \& Rees}{1993}]{Meszaros1993} 
Meszaros P., Rees M.~J., 1993, ApJ, 405, 278

\bibitem[\protect\citeauthoryear{M{\'e}sz{\'a}ros \& Rees}{1997}]{Meszaros1997} 
M{\'e}sz{\'a}ros P., Rees M.~J., 1997, ApJ, 476, 232

\bibitem[\protect\citeauthoryear{Miceli \& Nava}{2022}]{Miceli2022}
Miceli D., Nava L., 2022, Galax, 10, 66

\bibitem[\protect\citeauthoryear{Murase et al.}{2010}]{Murase2010}
Murase K., Toma K., Yamazaki R., Nagataki S., Ioka K., 2010, MNRAS, 402, L54
 
\bibitem[\protect\citeauthoryear{Murase et al.}{2011}]{Murase2011} 
Murase K., Toma K., Yamazaki R., M{\'e}sz{\'a}ros P., 2011, ApJ, 732, 77

\bibitem[\protect\citeauthoryear{Nakar, Piran, \& Granot}{2002}]{Nakar2002} 
Nakar E., Piran T., Granot J., 2002, ApJ, 579, 699

\bibitem[\protect\citeauthoryear{Nakar, Ando, \& Sari}{2009}]{Nakar2009} 
Nakar E., Ando S., Sari R., 2009, ApJ, 703, 675

\bibitem[\protect\citeauthoryear{Nava et al.}{2013}]{Nava2013} 
Nava L., Sironi L., Ghisellini G., Celotti A., Ghirlanda G., 2013, MNRAS, 433, 2107

\bibitem[\protect\citeauthoryear{Pe'er}{2015}]{Peer2015} 
Pe'er A., 2015, AdAst, 2015, 907321

\bibitem[\protect\citeauthoryear{Piran}{1999}]{Piran1999} 
Piran T., 1999, PhR, 314, 575

\bibitem[\protect\citeauthoryear{Piran}{2004}]{Piran2004} 
Piran T., 2004, RvMP, 76, 1143

\bibitem[\protect\citeauthoryear{Ravasio et al.}{2019}]{Ravasio2019} 
Ravasio M.~E., et al., 2019, A\&A, 626, A12

\bibitem[\protect\citeauthoryear{Rhodes et al.}{2020}]{Rhodes2020} 
Rhodes L., et al., 2020, MNRAS, 496, 3326

\bibitem[\protect\citeauthoryear{Rhodes et al.}{2022}]{Rhodes2022} 
Rhodes L., van der Horst A.~J., Fender R., Aguilera-Dena D.~R., Bright J.~S., Vergani S., Williams D.~R.~A., 2022, MNRAS, 513, 1895

\bibitem[\protect\citeauthoryear{Ronchi et al.}{2020}]{Ronchi2020} 
Ronchi M., 2020, A\&A, 636, A55

\bibitem[\protect\citeauthoryear{Rossi et al.}{2004}]{Rossi2004} 
Rossi E.~M., Lazzati D., Salmonson J.~D., Ghisellini G., 2004, MNRAS, 354, 86

\bibitem[\protect\citeauthoryear{Sahu \& Fort{\'\i}n}{2020}]{Sahu2020} 
Sahu S., Fort{\'\i}n C.~E.~L., 2020, ApJL, 895, L41


\bibitem[\protect\citeauthoryear{Sahu, Valadez Polanco, \& Rajpoot}{2022}]{Sahu2022} 
Sahu S., Valadez Polanco I.~A., Rajpoot S., 2022, ApJ, 929, 70

%\bibitem[\protect\citeauthoryear{Sakamoto et al.}{2008}]{Sakamoto2008} 
%Sakamoto T., et al., 2008, ApJ, 679, 570

\bibitem[\protect\citeauthoryear{Salafia et al.}{2022}]{Salafia2021} 
Salafia O.~S., et al., 2022, ApJL, 931, L19

\bibitem[\protect\citeauthoryear{Sari, Piran \& Narayan}{1998}]{sari1998} 
Sari R., Piran T., Narayan R., 1998, ApJL, 497, L17

\bibitem[\protect\citeauthoryear{Sari}{1997}]{Sari1997} 
Sari R., 1997, ApJL, 489, L37

\bibitem[\protect\citeauthoryear{Sari \& Esin}{2001}]{Sari2001} 
Sari R., Esin A.~A., 2001, ApJ, 548, 787

\bibitem[\protect\citeauthoryear{Sato et al.}{2021}]{Sato2021} 
Sato Y., Obayashi K., Yamazaki R., Murase K., Ohira Y., 2021, MNRAS, 504, 5647

\bibitem[\protect\citeauthoryear{Selina et al.}{2018}]{Selina2018} 
Selina R.~J., et al., 2018, SPIE, 10700, 107001O

\bibitem[\protect\citeauthoryear{Shimoda \& Toma}{2021}]{Shimoda2021} 
Shimoda J., Toma K., 2021, ApJ, 913, 58

\bibitem[\protect\citeauthoryear{Shrestha et al.}{2022}]{Shrestha2022} 
Shrestha M., 2022, preprint (arXiv:2208.01729)

\bibitem[\protect\citeauthoryear{Totani \& Panaitescu}{2002}]{Totani2002} 
Totani T., Panaitescu A., 2002, ApJ, 576, 120

\bibitem[\protect\citeauthoryear{Tsuboi et al.}{2022}]{Tsuboi2022} 
Tsuboi M., Tsutsumi T., Miyazaki A., Miyawaki R., Miyoshi M., 2022, preprint (arXiv:2204.06778)

\bibitem[\protect\citeauthoryear{Urata et al.}{2015}]{Urata2015}
Urata Y., Huang K., Yamazaki R., Sakamoto T., 2015, ApJ, 806, 222

\bibitem[\protect\citeauthoryear{Urrutia, De Colle, \& L{\'o}pez-C{\'a}mara}{2022}]{Urrutia2022} 
Urrutia G., De Colle F., L{\'o}pez-C{\'a}mara D., 2022, preprint (arXiv:2207.07925)


\bibitem[\protect\citeauthoryear{Valeev et al.}{2019}]{Valeev2019} 
Valeev, A.F. et al. 2019, ATel, 25565, 1

\bibitem[\protect\citeauthoryear{Wang et al.}{2010}]{Wang2010} 
Wang X.-Y., He H.-N., Li Z., Wu X.-F., Dai Z.-G., 2010, ApJ, 712, 1232

\bibitem[\protect\citeauthoryear{Wang et al.}{2019}]{Wang2019} 
Wang X.-Y., Liu R.-Y., Zhang H.-M., Xi S.-Q., Zhang B., 2019, ApJ, 884, 117

\bibitem[\protect\citeauthoryear{Yamasaki \& Piran}{2022}]{Yamasaki2022} 
Yamasaki S., Piran T., 2022, MNRAS, 512, 2142

\bibitem[\protect\citeauthoryear{Yamazaki, Ioka, \& Nakamura}{2002}]{Yamazaki2002} 
Yamazaki R., Ioka K., Nakamura T., 2002, ApJL, 571, L31

\bibitem[\protect\citeauthoryear{Yamazaki, Ioka, \& Nakamura}{2003}]{Yamazaki2003a} 
Yamazaki R., Ioka K., Nakamura T., 2003, ApJ, 593, 941

\bibitem[\protect\citeauthoryear{Yamazaki, Yonetoku, \& Nakamura}{2003}]{Yamazaki2003b} 
Yamazaki R., Yonetoku D., Nakamura T., 2003, ApJL, 594, L79

\bibitem[\protect\citeauthoryear{Yamazaki, Ioka, \& Nakamura}{2004}]{Yamazaki2004} 
Yamazaki R., Ioka K., Nakamura T., 2004, ApJL, 606, L33

\bibitem[\protect\citeauthoryear{Yamazaki et al.}{2020}]{yamazaki2020}
Yamazaki R., Sato Y., Sakamoto T., Serino M., 2020, MNRAS, 494, 5259

\bibitem[\protect\citeauthoryear{Zhang}{2018}]{Zhang2018} 
Zhang B., 2018, {\it The Physics of Gamma-ray Bursts}, Cambridge Univ. Press, Cambridge, doi:10.1017/9781139226530

\bibitem[\protect\citeauthoryear{Zhang et al.}{2021b}]{BTZhang2021} 
Zhang B.~T., Murase K., Veres P., M{\'e}sz{\'a}ros P. 2021b, ApJ, 920, 55

\bibitem[\protect\citeauthoryear{Zhang et al.}{2021a}]{LLZhang2021} 
Zhang L.-L., Ren J., Huang X.-L., Liang Y.-F., Lin D.-B., Liang E.-W., 2021a, ApJ, 917, 95

\bibitem[\protect\citeauthoryear{Zhang \& MacFadyen}{2009}]{Zhang2009} 
Zhang W., MacFadyen A., 2009, ApJ, 698, 1261



\end{thebibliography}

\end{document}